\documentclass[11pt,a4paper]{article}
\pdfoutput = 1
\usepackage{jcappub}
\usepackage[utf8]{inputenc}
\usepackage[T1]{fontenc}
\usepackage{amsmath,amssymb,bm}
\usepackage{graphicx}
\usepackage{hyperref}
\usepackage{natbib}
\usepackage{caption}
\usepackage{subcaption}
\usepackage{tablefootnote}
\usepackage{soul}

\renewcommand{\vec}[1]{\bm{#1}}

\newcommand*{\df}  {\delta}

\newcommand*{\non}  {\nonumber}
\newcommand*{\lb}  {\left(}
\newcommand*{\rb}  {\right)}

\newcommand*{\la}  {\left\langle}
\newcommand*{\ra}  {\right\rangle}
\newcommand*{\eps}  {\epsilon}

\newcommand{\ba}{\[\begin{aligned}}
\newcommand{\ea}{\end{aligned}\]}
\newcommand{\eq}[1]{\begin{align}#1\end{align}}
\newcommand{\eeq}[1]{\begin{equation}#1\end{equation}}

\subheader{
\footnotesize
\vspace*{-3.7em}
\begin{flushright}
LAPTH-026/26 \\
RBI-ThPhys-2025-14
\end{flushright}
}

\author[a]{Zucheng Gao}
\author[d,c,b]{Zvonimir Vlah}
\author[b,c,e]{Anthony Challinor}

\affiliation[a]{Laboratoire d'Annecy de
Physique Théorique, CNRS, 9 Chemin de Bellevue, 74940 Annecy, France.}
\affiliation[b]{Institute of Astronomy, Madingley Road, Cambridge, CB3 0HA, UK,}
\affiliation[c]{Kavli Institute for Cosmology Cambridge, Madingley Road, Cambridge, CB3 0HA, UK,}
\affiliation[d]{Division of Theoretical Physics, 
Ru\dj er Bo\v{s}kovi\'{c} Institute, Zagreb HR-10000, Croatia,}
\affiliation[e]{DAMTP, Centre for Mathematical Sciences, Wilberforce Road, Cambridge CB3 0WA, UK,}

\emailAdd{zg285@cantab.ac.uk}
\emailAdd{zvlah@irb.hr}
\emailAdd{a.d.challinor@ast.cam.ac.uk}

\title{Efficient computation of the galaxy angular bispectrum in redshift space}

\keywords{bispectrum, galaxy clustering, redshift space distortions}

\abstract{
Efficient computation of the angular bispectrum is an essential part of modelling large-scale structure observations, but it still remains an extremely challenging task. In this work, we compute the tree-level, unequal-time angular bispectrum in both real and redshift space. By deriving full-sky results, we show that the bispectrum can be expressed as a sum of products of two angular power spectra, enabling the use of our recently developed flat-sky approximation to enhance computational efficiency significantly. This flat-sky formalism preserves key line-of-sight mode information while discarding extraneous full-sky contributions. We validate our approach by comparing it with direct full-sky integration, finding excellent agreement across a wide range of scales and redshifts for all bispectrum configurations. At redshift $z = 1$, we achieve sub-percent agreement (for multipoles $\ell \gtrsim 5$) between full-sky and flat-sky results for equilateral, squeezed, and folded configurations, using narrow Gaussian radial window functions ($\sigma_z = 0.01$) in both equal-time and unequal-time scenarios. On small scales, where direct full-sky integration becomes computationally prohibitive, our results align with the Limber approximation (where applicable), confirming the robustness and accuracy of our implementation. To facilitate future studies, we provide a \texttt{Python} implementation of our results, which is publicly available on \texttt{GitHub}. 
}

\begin{document}

\maketitle

\section{Introduction}
\label{Sec:Intro}

With powerful new galaxy surveys with instruments such as Euclid~\cite{Euclid2011} and DESI~\cite{DESI2016} underway, and the imminent Rubin Observatory~\cite{LSSTDarkEnergyScience:2022lno} and launch of the Roman Telescope~\cite{Roman2015}, we have entered a new era of precision cosmology. These surveys have large sky coverage and high survey depth, resulting in a high number of detected galaxies. To extract cosmological information reliably from these large datasets, the theoretical modeling needs to be highly accurate and computationally efficient. 

The three-point correlation function, or bispectrum (its Fourier counterpart), is an important quantity to study primordial non-Gaussianity and other potential additions to the standard $\Lambda$CDM model, such as modified gravity. Even within the standard model, the bispectrum can play an important role in breaking degeneracies between the growth of structure and galaxy bias. Including redshift space distortions (RSD) is essential in modelling the bispectrum since their effect can partly mimic other large-scale signals of interest, such as local-model primordial non-Gaussianity. The 3D bispectrum (in Cartesian Fourier coordinates) with RSD included has been investigated intensively in the literature \cite{Gil-Marin:2014,Gil-Marin:2016,Cabass:2022,Ivanov:2023,Desjacques:2018}. These calculations usually assume the plane-parallel (flat-sky) approximation, which makes the subsequent addition of wide-angle effects a daunting task \cite{Pardede:2023, Benabou:2023}. 

Alternatively, the full-sky angular bispectrum (which is directly observable) intrinsically includes these projection effects, and can be naturally extended to include the full set of relativistic effects~\cite{Yoo:2014,DiDio:2015, Montandon:2025uul}. However, the direct computation of the angular bispectrum is a computationally challenging task, so it is seldom used in practical data analysis. Even in its simplest form, considering only second-order biasing, modelling the projected full-sky angular bispectrum is computationally demanding. To address this, reference \cite{Assassi2017} introduced an FFTLog-based approach that leverages the separability of the tree-level bispectrum to accelerate computations. Despite this improvement, the method still depends on complex hypergeometric functions, which are expensive to evaluate. Subsequently, reference \cite{DiDio_2018} extended this work to redshift space, while reference \cite{Montandon:2025uul} incorporated a set of the relativistic effects. In practice, additional integration of survey-specific radial window functions is needed to obtain the observed angular bispectrum, making this process too slow for a practical Markov Chain Monte Carlo (MCMC) analysis. To overcome these difficulties, one may turn to machine learning interpolation and emulation techniques, though these inevitably come with additional drawbacks, like lack of physical insights and lead to a lack of interpretability. We focus on alternative strategies based on analytical and computational simplifications. 

In our previous work~\cite{Gao_2024}, we showed that the flat-sky result (derived as a leading contribution in the asymptotic expansion) can approximate the full-sky angular power spectrum (two-point correlation function) to a percent-level even on the largest scales. At the same time, it can be computed numerically very efficiently, making it especially suitable for the MCMC analysis. In this work, we apply our flat-sky approximation to the angular bispectrum. We first (re)derive the general form for the tree-level full-sky angular bispectrum with RSD, starting from the observed galaxy number counts in redshift space at second order in perturbation theory, which also includes wide-angle effects and non-distant-observer corrections.
We start directly from the full-sky angular bispectrum and approximate integrals over wavenumber, which involve products of spherical Bessel functions, with their flat-sky Fourier equivalents by generalising results from Gao et al.~\cite{Gao_2023}. This method naturally results in a set of correction terms, which improve the accuracy in approximating the full-sky angular bispectrum. In other words, by using the correspondence between full- and flat-sky integrals on the general result for the full-sky angular bispectrum, we naturally obtain correction terms for wide-angle effects.

The paper is organised as follows: in section~\ref{sec:fullsky_real}, we first present our real-space results for the unequal-time angular bispectrum. We extend these to redshift space in section~\ref{sec:full_bispec_rsd}. We conclude in section~\ref{sec:conclusion}. In appendix~\ref{append:misalign}, we check the validity of the distant-observer approximation. 

We work with the \textit{Planck}, flat, $\Lambda\text{CDM}$ cosmology \cite{Aghanim:2018}, with CDM density $\Omega_c h^2 = 0.11933$, baryon density $\Omega_b h^2 = 0.02242$, Hubble constant $H_0 = 100 h \, \text{km\,s}^{-1}\,\text{Mpc}^{-1}$ with $h=0.6766$, scalar spectral index $n_s=0.9665$, and fluctuation amplitude $\sigma_8 = 0.8103$. All results presented here were obtained using the \texttt{Python} code that we made publicly available on \texttt{GitHub}.\footnote{https://github.com/GZCPhysics/Bispectrum.git} 

\section{Angular bispectrum in real space}
\label{sec:fullsky_real}

In this section, we focus on the real-space angular bispectrum of a biased tracer at leading order (tree-level) in cosmological perturbation theory. We outline the derivation of the full-sky angular bispectrum and detail the numerical optimisation techniques employed in its computation. Additionally, we compare with the Limber approximation (while we discuss an extended version specifically for the squeezed limit in appendix~\ref{append:extended_limber}). 
Finally, we present numerical results for the real-space bispectrum, highlighting key features and implications.

\subsection{Real space overdensity and the angular bispectrum}
\label{subsec:bispec_real_generic}

To obtain the tree-level bispectrum results, let us first consider the expansion of the overdensity field of a biased tracer up to the second-order:
\eq{\label{eq:delta_expansion_in_real_space}
\delta(\vec{x};\; z) = \delta^{(1)}(\vec{x};\; z) + \delta^{(2)} (\vec{x};\; z) + \cdots \, ,
}
where each order can be related to the linear matter overdensity field, $\delta_{\rm G}$ (the subscript ``G'' indicates that the field is Gaussian). In Fourier space, we have
\eq{
\label{eq:delta_1_real_space}
\delta^{(1)} (\vec{k};\; z) &= b_{10}(z)\delta_{\mathrm{G}}(\vec{k};\; z)\, , \\
\label{eq:delta_2_real_space}
\delta^{(2)} (\vec{k};\; z) &=\int \frac{d^3 \vec{k}_1}{(2\pi)^3}\frac{d^3 \vec{k}_{2}}{(2\pi)^3} (2\pi)^3 \delta^{\rm D}(\vec k  - \vec k_1 - \vec k_2) \mathcal{K}_2(\vec{k}_1, \vec{k}_2;\, z) \delta_{\mathrm{G}}(\vec k_1;\; z) \delta_{\mathrm{G}}(\vec k_2;\; z) \, ,
}
where we introduced the second-order tracer kernel 
\eeq{
\mathcal{K}_2(\vec{k}_1, \vec{k}_2;\, z) = b_{10}(z) \mathcal{F}_2(\vec k_1,\vec k_2) + b_{20}(z) + b_{K^2}(z)\mathcal{S}_2(\vec k_1,\vec k_2) \, .
}
Alongside the quadratic bias term ($b_{20}$), this contains the second-order matter kernel $\mathcal{F}_2$ and shear kernel $\mathcal{S}_2$, explicitly given by
\eq{
\mathcal{F}_2(\vec k_1,\vec k_2) &= \frac{5}{7} + \frac{1}{2}\frac{\vec k_1 \cdot \vec k_2}{k_1 k_2} \lb\frac{k_1}{k_2} + \frac{k_2}{k_1}  \rb + \frac{2}{7} \lb \frac{\vec k_1 \cdot \vec k_2}{k_1 k_2}\rb^2\non \\
\label{eq:F_kernel}
& = \frac{5}{14} - \frac{5}{28} \frac{k_1^2}{k_2^2} - \frac{5}{28} \frac{k_2^2}{k_1^2} + \frac{3}{28} \frac{k^2}{k_1^2} + \frac{3}{28} \frac{k^2}{k_2^2} + \frac{1}{14}\frac{k^4}{k_1^2 k_2^2} \, , \\
\mathcal{S}_2(\vec k_1, \vec k_2) &= \frac{(\vec k_1 \cdot \vec k_2)^2}{k_1^2 k_2^2} - \frac{1}{3} \nonumber\\
&= \frac{1}{6} + \frac{1}{4} \frac{k_1^2}{k_2^2} + \frac{1}{4} \frac{k_2^2}{k_1^2} - \frac{1}{2} \frac{k^2}{k_2^2} - \frac{1}{2} \frac{k^2}{k_1^2} + \frac{1}{4} \frac{k^4}{k_1^2 k_2^2}\, , 
\label{eq:tidal_kernel}
}
where we have implicitly used the triangle relation, $\vec{k} = \vec{k}_1 + \vec{k}_2$, to write the kernels in a separable form. Note that we have made the quasi-EdS approximation for the time-dependence of the second-order matter density in writing down a time-independent matter kernel $\mathcal{F}_2$. In addition to the deterministic bias contributions, we can also add the two stochastic operators $\epsilon$ (both first- and second- order) and $\epsilon_\delta^{(1)} \delta_{\rm G}$ (second-order), which contribute up to the second order in the field \cite{Angulo:2015, Fujita:2016, Desjacques:2018}.

We consider the projection of an overdensity field on the sky
\eq{
\label{eq:delta_projection}
\delta(\hat{\vec{n}}) = \int d\chi\, W(\chi) \delta(\vec{x};\;\eta_0-\chi)\equiv \int_\chi \delta(\vec{x};\;\eta_0-\chi),
}
where $W(\chi)$ is the radial window function (selection function), $\eta_0$ is the conformal time of the present universe and $\chi$ is the look-back time, which has a one-to-one relation with the redshift. We note that $|\vec{x}|=\chi$. We also introduce a simplified notation $\int_\chi$ to indicate the integration over the radial window function.

Spherical-harmonic decomposition on the projected fields gives field multipole coefficients  
\eq{
\label{eq:general_decomposition}
\delta_{\ell m} = \int d^2 \hat{\vec{n}}\; \delta(\hat{\vec{n}}) Y^{*}_{\ell m}(\hat{\vec{n}}) \, ,
}
which for the first-order real-space overdensity, eq.~\eqref{eq:delta_1_real_space}, reads
\eq{
\label{eq:delta_1_decomposition_real}
\delta^{(1)}_{\ell m} 
&= \int_\chi \int d^2 \hat{\vec{x}}\, Y_{\ell m}^{*}\lb \hat{\vec{x}} \rb  \int \frac{d^3 \vec{k}}{(2\pi)^3}\,  e^{i\vec{k}\cdot\vec{x}} \delta^{(1)}(\vec{k};\; z[\chi]) \non\\
& = 4\pi i^{\ell}\int_\chi \int \frac{d^3 \vec{k} }{(2\pi)^3} j_{\ell}\lb k\chi \rb \delta^{(1)}(\vec{k};\; z[\chi]) 
Y_{\ell m}^{*}(\hat{\vec{k}}) \, ,
}
where we applied the plane-wave expansion. 
Similarly, we calculate the harmonic expansion of the second-order field, eq.~\eqref{eq:delta_2_real_space}: 
\eq{
\label{eq:delta_2_decomposition_real}
\delta^{(2)}_{\ell m} 
&= \int_\chi \int d^2 \hat{\vec{x}}\, Y_{\ell m}^{*}\lb \hat{\vec{x}} \rb \int \frac{d^3 \vec{k} }{(2\pi)^3} e^{i\vec{k}\cdot\vec{x}} \delta^{(2)}(\vec{k};\; z[\chi]) \non \\
& = (4\pi)^4 \sum_{\ell_1 m_1} \sum_{\ell_2 m_2} i^{\ell_1 + \ell_2} \mathcal{G}_{\ell \ell_1 \ell_2}^{m m_1 m_2}\int_\chi \int \frac{k^2 dk }{(2\pi)^3} j_{\ell}\lb k\chi\rb \int r^2 dr \int \frac{d^3 \vec{k}_1}{(2\pi)^3} \frac{d^3 \vec{k}_2}{(2\pi)^3}\, j_{\ell}\lb k r\rb j_{\ell_1}\lb k_1 r\rb j_{\ell_2}\lb k_2 r\rb  \non \\
&\hspace{3.4cm}\times
\mathcal{K}_2(\vec{k}_1,\vec{k}_2;\; z[\chi]) \delta_{\mathrm{G}}(\vec{k}_1;\; z[\chi]) \delta_{\mathrm{G}}(\vec{k}_2;\; z[\chi])
 Y_{\ell_1 m_1}(\hat{\vec{k}}_1) Y_{\ell_2 m_2}(\hat{\vec{k}}_2)\, ,
}
where we have introduced the (real-valued) Gaunt coefficient 
\eq{\label{eq:Gaunts_coefficient}
\mathcal{G}_{\ell_1 \ell_2 \ell_3}^{m_1 m_2 m_3} &= \int d^2 \hat{\vec{n}}\, Y_{\ell_1 m_1}\lb \hat{\vec{n}}\rb Y_{\ell_2 m_2}\lb \hat{\vec{n}}\rb Y_{\ell_3 m_3}\lb \hat{\vec{n}}\rb \non \\
& =  \sqrt{\frac{(2\ell_{1}+1)(2\ell_{2}+1)(2\ell_{3}+1)}{4\pi}}\begin{pmatrix}\ell_{1}&\ell_{2}&\ell_{3}\\0&0&0\end{pmatrix}\begin{pmatrix}\ell_{1}&\ell_{2}&\ell_{3}\\m_{1}&m_{2}&m_{3}\end{pmatrix}\, .
}

Once we have an explicit form of the projected fields, we are ready to compute the angular correlations.
The leading-order (tree-level) angular three-point correlator is then given by
\eq{
\label{eq:B_real}
\la \delta_{\ell_1 m_1}  \delta_{\ell_2 m_2}  \delta_{\ell_3 m_3} \ra 
&= \la \delta_{\ell_1 m_1}^{(1)}  \delta_{\ell_2 m_2}^{(1)}  \delta_{\ell_3 m_3}^{(2)} \ra + 2\, \mathrm{perms.} \\
&= \lb \frac{2}{\pi}\rb^3 \mathcal{G}_{\ell_1 \ell_2 \ell_3}^{m_1 m_2 m_3 } \int_{\chi_1 \chi_2 \chi_3}  \int r^2 d r \prod_{i=1}^3 \int k_i^2 d k_i j_{\ell_i} (k_i r)  j_{\ell_i} (k_i \chi_i) \non\\
&\hspace{1.2cm} \times 2\mathcal{K}_2(\vec{k}_1,\vec{k}_2;\; z_3) b_{10}(z_1) b_{10}(z_2) P_L(k_1;\; z_1, z_3) P_L(k_2;\; z_2, z_3) + 2\, \mathrm{perms.} \non \\
&\hspace{10.5cm} +\,  \mathrm{stochastic} \, , \non
}
where we give the explicit form of the stochastic terms later on in eq.~\eqref{eq:Bis_tree_real}. We note that $\mathcal{K}_2(\vec{k}_1,\vec{k}_2)$ should be thought of as a function of $k_1$, $k_2$ and $k_3$ in this expression, with $\vec{k}_3 = -(\vec{k}_1+\vec{k}_2)$. Defining the (reduced) angular bispectrum as
\eq{
\la \delta_{\ell_1 m_1}  \delta_{\ell_2 m_2}  \delta_{\ell_3 m_3} \ra = \mathcal G^{m_1 m_2 m_3}_{\ell_1 \ell_2 \ell_3} B_{\ell_1 \ell_2 \ell_3} \, ,
\label{eq:bis_real_final_step}
}
and introducing the \textit{unequal-time} angular bispectrum 
\eeq{
\label{eq:bispec_unprojected_bispec}
B_{\ell_1 \ell_2 \ell_3}
=  \int_{\chi_1\chi_2\chi_3} \mathbb B_{\ell_1 \ell_2 \ell_3} (\chi_1, \chi_2, \chi_3)\, ,
}
we can explicitly write 
\eq{
\label{eq:reduced_angle_bispec}
\mathbb B_{\ell_1 \ell_2 \ell_3} (\chi_1, \chi_2, \chi_3)
= \lb \frac{2}{\pi} \rb^3 \int \lb \prod_{i=1}^3 k_i^2 d k_i j_{\ell_i} ( k_i \chi_i) \rb
& \mathcal D_{\ell_1 \ell_2 \ell_3}(k_1,k_2,k_3) \non\\
& \times B\lb k_1,k_2,k_3; z[\chi_1], z[\chi_2], z[\chi_3] \rb\, ,
}
thus connecting the angular bispectrum and the 3D bispectrum.
In the expression above we use a geometrical projection function
\eeq{
\label{eq:projection_factor}
\mathcal D_{\ell_1 \ell_2 \ell_3}(k_1,k_2,k_3) = \int r^2 d r \, j_{\ell_1} (k_1 r)  j_{\ell_2} (k_2 r)  j_{\ell_3} (k_3 r)  \, ,
}
which projects the unequal-time 3D bispectrum to the angular bispectrum. We
also used the explicit tree-level form of the unequal-time 3D bispectrum
\eq{
\label{eq:Bis_tree_real}
B\lb k_1,k_2,k_3; z_1, z_2, z_3 \rb
=2\mathcal{K}_2&(\vec{k}_1,\vec{k}_2; \; z_3) b_{10}(z_1) b_{10}(z_2) P_L(k_1;\, z_1, z_3) P_L(k_2;\, z_2, z_3) + 2\, \mathrm{perms.} \non\\
&+\, b_{10}(z_1) P_L(k_1;\, z_1, z_3)  P_{\eps \eps_\delta}(z_2, z_3)
+ 5\, \mathrm{perms.} \non\\
&+\, B_{\eps}(z_1,z_2,z_3)
\,,
}
even though it is straightforward to show that eq.~\eqref{eq:reduced_angle_bispec} is valid in general and does not rely on the tree-level perturbative solution. The second and the third lines in eq.~\eqref{eq:Bis_tree_real} arise from the stochastic contributions in the biased tracer field and involve the cross-power spectrum between $\epsilon^{(1)}$ and $\epsilon_\delta^{(1)}$ and the 3D bispectrum of $\epsilon$, respectively, which are both assumed to be independent of $k$. Note also that the angular bispectrum, as given in eq.~\eqref{eq:bispec_unprojected_bispec}, reduces to the unequal-time angular bispectrum $\mathbb B_{\ell_1 \ell_2 \ell_3}$ in the limit of narrow window functions $W(\chi)$, i.e., when the integral over the window functions becomes trivial.

\subsection{Tree-level results}
\label{subsec:tree_level_bispec_real}

From now on, we use the linear theory of the 3D power spectrum, where we can separate the scale and time dependence $P_L(k;\, z_1, z_2) = D(z_1) D(z_2) P_L(k)$, where $D(z)$ is the linear growth factor and $P_L(k)$ the linear matter power spectrum at $z=0$. In fact, the unequal-time power spectrum de-correlates much faster than linear theory suggests, as the long displacement dispersion dampens short scales~\cite{Raccanelli+:2023}. 
Our angular bispectrum thus reduces to the following form 
\eq{
\label{eq:bispec_real_explicit}
B_{\ell_1\ell_2\ell_3}
& = \lb \frac{2}{\pi}\rb^3 \int_{\chi_1 \chi_2 \chi_3} D_1 D_2 D_3^2\,  \prod_{i=1}^{3} \int k_i^2 dk_i j_{\ell_i} (k_i \chi_i) \int r^2 dr\, j_{\ell_1}\lb k_1 r\rb j_{\ell_2}\lb k_2 r\rb  j_{\ell_3}\lb k_3 r\rb \non \\
&\hspace{2.0cm} \times 2 \mathcal{K}_2(\vec{k}_1,\vec{k}_2;\; z_3) b_{10}(z_1) b_{10}(z_2) P_L(k_1) P_L(k_2) + 2\, \mathrm{perms.}
+ \mathrm{stochastic}\, ,
}
where we include the permutations among three indices, and we also define the linear growth factor $D_i \equiv D(z[\chi_i])$. We also take the redshift evolution of bias into account. 

The unequal-time tree-level angular bispectrum can thus be written as the linear combination of terms
\eq{
\label{eq:Blll_nnn_real}
\mathbb B_{\ell_1\ell_2\ell_3}^{n_1 n_2 n_3}  (\chi_1,\chi_2,\chi_3) \equiv 
2 D_1 D_2 D_3^2 \int r^2 dr\,\mathbb{C}_{n_1}(\ell_1, r, \chi_1) \mathbb{C}_{n_2}(\ell_2, r, \chi_2) \mathbb{I}_{n_3}(\ell_3, r, \chi_3) \, ,
}
where we introduced 
\eq{
\label{eq:unprojected_Cn_real}
\mathbb{C}_{n}(\ell, r, \chi) &\equiv  \frac{2}{\pi} \int k^2 dk\, j_{\ell}(kr)j_{\ell}(k\chi) k^{2n} P_L(k)\, , \\
\label{eq:unprojected_In_real}
\mathbb{I}_{n}(\ell, r, \chi) &\equiv  \frac{2}{\pi} \int k^2 dk\, j_{\ell}(kr)j_{\ell}(k\chi) k^{2n}\, .
}
Combining all the terms, the tree-level angular bispectrum for the $\mathcal{K}_2$ kernel is then given by
\eq{
\label{eq:combination}
& \mathbb B_{\ell_1\ell_2\ell_3} (\chi_1,\chi_2,\chi_3) \non\\
&\hspace{0.2cm} = b_{10}(\chi_1)b_{10}(\chi_2)\lb \tfrac{5}{7}b_{10} + b_{20} + \tfrac{1}{3}b_{K^2} \rb(\chi_3) \mathbb B_{\ell_1\ell_2\ell_3}^{000} + b_{10}(\chi_1)b_{10}(\chi_2)\lb -\tfrac{5}{14}b_{10} + \tfrac{1}{2}b_{K^2} \rb (\chi_3) \mathbb B_{\ell_1\ell_2\ell_3}^{1-10} \non\\
&\hspace{0.5cm} + b_{10}(\chi_1)b_{10}(\chi_2)\lb -\tfrac{5}{14}b_{10} + \tfrac{1}{2}b_{K^2} \rb (\chi_3) \mathbb B_{\ell_1\ell_2\ell_3}^{-110}+ b_{10}(\chi_1)b_{10}(\chi_2)\lb \tfrac{3}{14}b_{10}-b_{K^2} \rb (\chi_3) \mathbb B_{\ell_1\ell_2\ell_3}^{-101} \non\\
&\hspace{0.5cm} + b_{10}(\chi_1)b_{10}(\chi_2)\lb \tfrac{3}{14}b_{10}-b_{K^2} \rb (\chi_3) \mathbb B_{\ell_1\ell_2\ell_3}^{0-11} + b_{10}(\chi_1)b_{10}(\chi_2)\lb \tfrac{1}{7}b_{10}+\tfrac{1}{2}b_{K^2}  \rb (\chi_3) \mathbb B_{\ell_1\ell_2\ell_3}^{-1-12}  \non\\
&\hspace{0.5cm}+ 2\, \mathrm{perms.}
}
Notice that only $n=-1, 0, 1$ cases of the $\mathbb{C}_n$ terms contribute to the full bispectrum expression, while for $\mathbb{I}_n$ terms, we require $n=0,1,2$. To obtain the final projected bispectrum, one needs to do the final integral over the window functions, eq.~\eqref{eq:bis_real_final_step}. Up to this point, the results presented for the deterministic bias contributions are equivalent to the ones obtained in \cite{Assassi2017}, what differs is the method of computation of these terms, which we discuss in the next subsection. 

In addition to the deterministic bias terms, we also have stochastic contributions. However, their form is significantly simpler and we can write
\eq{
B_{\ell_1\ell_2\ell_3}^{\rm stoch.} = 
2 \int \frac{d\chi d\chi'}{\chi'^2}\,   W_1(\chi)W_2(\chi')W_3(\chi')\,  b_{10}(\chi) & D(\chi) D(\chi') P_{\eps \eps_\delta}(\chi', \chi') \mathbb{C}_{0}(\ell_1, \chi, \chi') 
+ 2\, \mathrm{perms.}\, \non\\
& + \int \frac{d\chi}{\chi^4}\, W_1(\chi)W_2(\chi)W_3(\chi)
B_{\eps}(\chi,\chi,\chi)\, ,
\label{eq:realspacestoch}
}
where we notice that the scale dependence of the first term comes solely from one $\mathbb{C}_{0}(\ell_1)$, while the second (purely stochastic) term exhibits no scale dependence, as expected.

\subsection{Numerical evaluation with flat-sky approximation}
\label{subsec:numerical_evaluation}

In our previous work~\cite{Gao_2023, Gao_2024}, we showed that $\mathbb{C}_{n}(\ell)$ terms in eq.~\eqref{eq:unprojected_Cn_real} could be approximated by a flat-sky expression (to sub-percent level), thus providing an efficient computational method. Here, we again apply the flat-sky approximation for the evaluation of the unprojected power spectra $\mathbb{C}_n(\ell)$. As a brief reminder, we approximate
\eq{
\label{eq:unproject_Cn_flat}
\mathbb{C}_{n}(\ell, r, \chi) \approx \frac{1}{r\chi} \int \frac{dk_\parallel}{2\pi} e^{ik_\parallel (\chi - r)} \lb k_\parallel^2 + \tilde{\ell}^2 \rb^n P_L\lb \sqrt{k_\parallel^2 + \tilde{\ell}^2} \rb\, , 
}
where we defined $\tilde{\ell}^2 \equiv \ell(\ell+1)/(r\chi)$. 
We can utilize the FFTLog expansion $P_L(k) = \sum_i \alpha_i k^{\nu_i}$ of the linear power spectrum, where the coefficients $\alpha_i$ carry all the dependence on cosmology, and the $\nu_i$ are complex powers. 
The unprojected angular power spectrum can then be evaluated as
\eq{
\label{eq:Expansion}
\mathbb{C}_{n}(\ell, r, \chi) = \sum_i \alpha_i \tilde{\ell}^{1+\nu_i+2n} M_{\nu_i+2n}^{(0)}(x) \, ,
}
where $x = \tilde{\ell}(\chi-r)$. Here,
\eq{\label{eq:M0i}
M_{\nu_i+2n}^{(0)}(x) \equiv \frac{2^{\frac{1}{2}(1+\nu_i)+n}}{\sqrt{\pi}\Gamma(-\frac{\nu_i}{2}-n)} x^{-\frac{1}{2}(1+\nu_i)-n} K_{\frac{1}{2}(1+\nu_i)+n}(x)\, ,
}
where $K_\nu(x)$ are modified Bessel functions of the second kind. One of the advantages of this approach is that the series of functions $M_{\nu}^{(0)}(x)$ can be precomputed and stored in memory, thus providing a convenient and rapid evaluation method. The superscript $(0)$ is for later convenience since the functions $M_{\nu}^{(m)}$ generalize later once we consider redshift space distortions. 

Care is needed for $n=1$. In this case, the (Basset) integral that gives rise to eq.~\eqref{eq:Expansion} requires $\mathrm{Re}(\nu_i)<-2$. However, this conflicts with the requirement $\mathrm{Re}(\nu_i)>-2$ for the FFTLog expansion to represent $P_L(k)$ accurately given its asymptotic behaviour $P_L(k) \sim k^{-3} \ln^2(k/k_{\text{eq}})$, where $k_{\text{eq}}$ is the equality scale. Indeed, we found the $n=1$ results to be numerically unstable. Thus, we do not directly use this representation for $\mathbb{C}_{n=1}$. Instead, we use the differential equation for the spherical Bessel function,
\eq{
\label{eq:derivative_spherical_bessel_function}
\left[ -\frac{\partial^2}{\partial r^2} -\frac{2}{r}\frac{\partial}{\partial r}+\frac{\ell(\ell+1)}{r^2} \right] j_{\ell}(kr) = k^2 j_{\ell}(kr) \, ,
}
in eq.~\eqref{eq:unprojected_Cn_real} to obtain the relation
\eq{\label{eq:Cn1}
\mathbb{C}_{n=1}(\ell, r, \chi) = \left[ -\frac{\partial^2}{\partial r^2} -\frac{2}{r}\frac{\partial}{\partial r}+\frac{\ell(\ell+1)}{r^2} \right] \mathbb{C}_{n=0}(\ell, r, \chi) \, .
}
This provides us with a representation of $\mathbb{C}_{1}$ via the differential operator acting on  $\mathbb{C}_{0}$ that can be evaluated using our flat-sky FFTLog-based methods above. We can then evaluate the differential operator either numerically or analytically.

Since there is no power spectrum involved in $\mathbb{I}_n$ terms, we can keep the full-sky form and the three terms are given as
\eq{
\label{eq:In_real}
\mathbb{I}_{n}\lb \ell, r, \chi\rb = \left[ -\frac{\partial^2}{\partial \chi^2} -\frac{2}{\chi}\frac{\partial}{\partial \chi}+\frac{\ell(\ell+1)}{\chi^2} \right]^{n} \left(\frac{1}{\chi^2} \delta^{\mathrm{D}}(\chi-r)\right),
}
with $n=0,1,2$. Note that this equation is symmetric in $r$ and $\chi$, and, thus, one is free to swap the variables in the differential operator. Written in terms of $\chi$, we can apply integration-by-parts on the window function and linear growth factor. For example, the integral of eq.~\eqref{eq:Blll_nnn_real} over the radial window functions becomes
\eq{\label{eq:Blll_nnn_real_with_In}
B_{\ell_1\ell_2\ell_3}^{n_1 n_2 n_3} &= 2\int d\chi_1 d\chi_2 d\chi_3 W_1D_1 W_2D_2 
\,
\left[ -\frac{\partial^2}{\partial \chi_3^2} +\frac{\partial}{\partial \chi_3}\frac{2}{\chi_3}+\frac{\ell_3(\ell_3+1)}{\chi_3^2} \right]^{n_3} \lb W_3D^2_3 \rb  \non \\
&\hspace{5cm}\times \int r^2 dr\, \mathbb{C}_{n_1}(\ell_1, r, \chi_1) \mathbb{C}_{n_2}(\ell_2, r, \chi_2) \frac{1}{\chi_3^2} \delta^{\mathrm{D}}(\chi_3-r) \non \\
& = 2\int d\chi_1 d\chi_2 d\chi_3 W_1D_1 W_2D_2
\,
\left[ -\frac{\partial^2}{\partial \chi_3^2} +\frac{\partial}{\partial \chi_3}\frac{2}{\chi_3}+\frac{\ell_3(\ell_3+1)}{\chi_3^2} \right]^{n_3} \lb W_3D^2_3 \rb \non \\
&\hspace{7cm}\times\mathbb{C}_{n_1}(\ell_1, \chi_3, \chi_1) \mathbb{C}_{n_2}(\ell_2, \chi_3, \chi_2)\, .
}
If, instead, we write the differential operator in terms of $r$, integration-by-parts acts on the product of two unprojected angular power spectra, which gives us
\eq{
\label{eq:Blll_nnn_real_with_In_derive_r}
B_{\ell_1\ell_2\ell_3}^{n_1 n_2 n_3} &= 2\int d\chi_1 d\chi_2 d\chi_3 W_1D_1 W_2D_2  W_3D^2_3  \non \\
&~~\times\left(\left[ -\frac{\partial^2}{\partial r^2} -\frac{2}{r}\frac{\partial}{\partial r} +\frac{\ell_3(\ell_3+1)}{r^2} \right]^{n_3} \mathbb{C}_{n_1}(\ell_1, r, \chi_1) \mathbb{C}_{n_2}(\ell_2, r, \chi_2) \right) _{r=\chi_3}\, ,
}
where have used the fact that the differential operator in eq.~\eqref{eq:derivative_spherical_bessel_function} is self-adjoint with respect to the weight function $r^2$.
In this work, we use eq.~\eqref{eq:Blll_nnn_real_with_In}, given that we are using (smooth) Gaussian window functions. However, in the case of general window functions, it may be preferable to use eq.~\eqref{eq:Blll_nnn_real_with_In_derive_r} instead. Note also that the boundary terms from integrating-by-parts can be non-zero when we consider lensing kernels in eq.~\eqref{eq:Blll_nnn_real_with_In}, but eq.~\eqref{eq:Blll_nnn_real_with_In_derive_r} works in all cases (requiring $\ell > 1$).

\subsection{Limber approximation}
\label{subsec:limber_real}

Care needs to be taken when we consider the Limber approximation for the angular bispectrum, especially for the squeezed limit. A simple, but not conceptually rigorous, way of effecting the Limber approximation is by setting
\eeq{
\label{eq:spherical_bessel_delta}
 j_{\ell} (k\chi)  \sim \sqrt{\frac{\pi}{2\ell}} \df^{\rm D} \lb k \chi - \ell \rb, \qquad \ell \to \infty
}
in eq.~\eqref{eq:bispec_real_explicit}. If we apply this approximation to all three multipoles we obtain the familiar Limber result (see, e.g., \cite{Troxel:2014})
\eeq{
\label{eq:bispec_real_explicit_limber}
B^{\rm Lim.}_{\ell_1\ell_2\ell_3}
= \int \frac{dr}{r^4}\, W_1(r) W_2(r) W_3(r) B\lb \frac{\ell_1}{r}, \frac{\ell_2}{r}, \frac{\ell_3}{r}; z[r], z[r], z[r]\rb.
}

We emphasize that this Limber approximation is only valid when all three multipoles are large, such that $1/\ell_i$ is small compared to the fractional width of the $i$th window function $W_i(r)$. This can be problematic for the squeezed limit of the bispectrum, where one mode is large scale. In particular, we find that one cannot then apply eq.~\eqref{eq:spherical_bessel_delta} to all three spherical Bessel functions in the geometrical projection function eq.~\eqref{eq:projection_factor}, while the approximation is also invalid for the spherical Bessel function associated with the soft mode. We provide alternative partial-Limber approximations for the squeezed limit in 
appendix~\ref{append:extended_limber}.

\subsection{Results}
\label{subsec:results_realspace}

\begin{figure}
    \centering
    \includegraphics[width=0.9\linewidth]{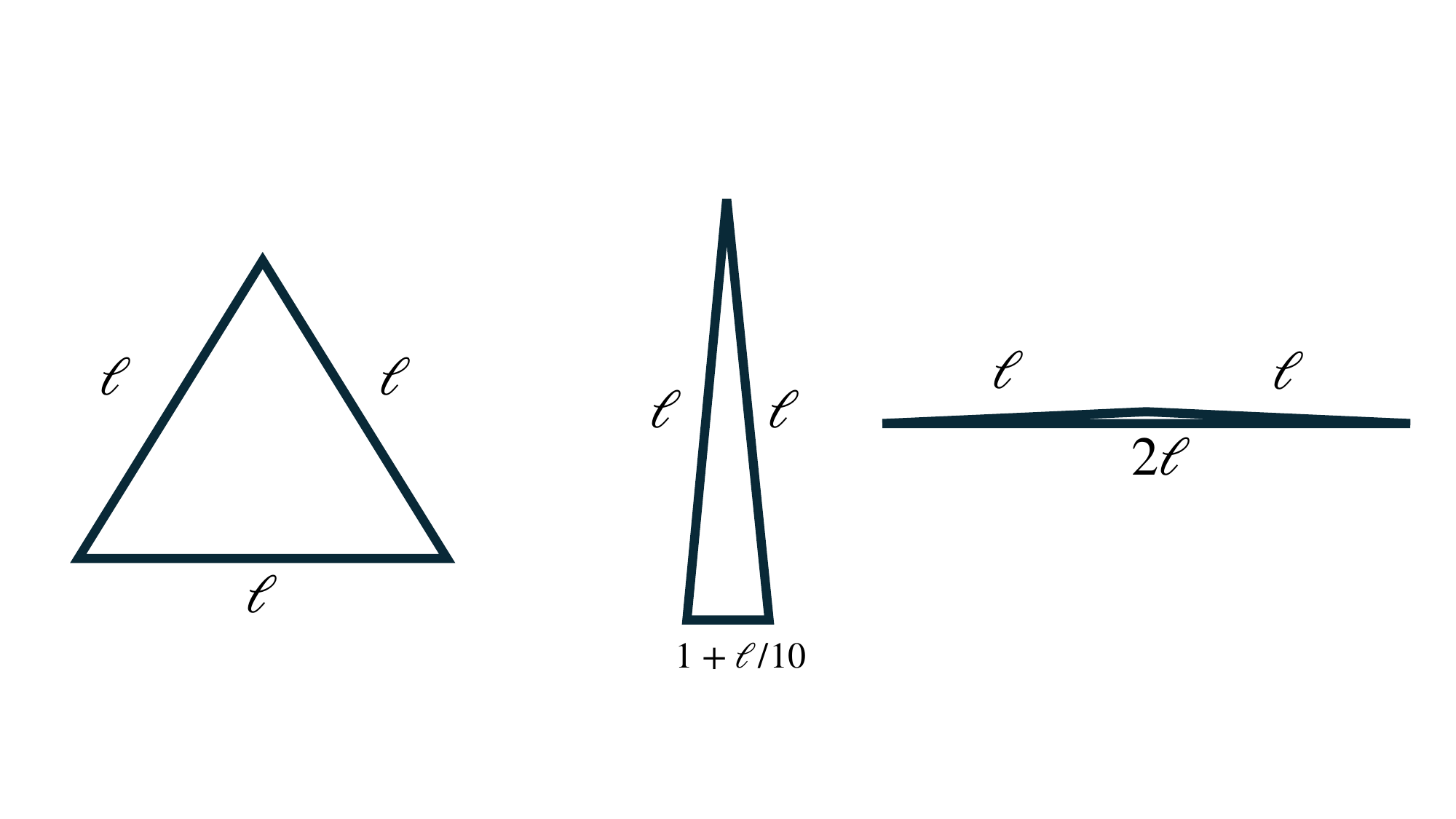}
    \caption{From left to right, we illustrate the equilateral, squeezed (with the soft multipole equal to $1+\ell/10$) and folded angular bispectrum configurations.}
    \label{fig:configuration}
\end{figure}

We now present numerical results for the real space angular bispectrum. We compare our results using the flat-sky approximations of section~\ref{subsec:numerical_evaluation} to brute-force computations of the full-sky results\footnote{We evaluate the unprojected angular power spectra $\mathbb{C}_n$, eq.~\eqref{eq:unprojected_Cn_real}, by direct numerical integration. The treatement of $\mathbb{I}_n$ is kept to be full-sky for both full-sky and flat-sky calculations since there is no power spectrum involved.}
and to the regular Limber approximation. We restrict our investigation to three triangle shapes: equilateral ($\ell_1 = \ell_2 = \ell_3 = \ell$); folded ($\ell_1 = \ell_2 = \ell_3/2 = \ell$); and squeezed ($\ell_1 = \ell_2 = \ell$, $\ell_3 = 1+\ell/10$). 
These cover the most diverse configuration choices, and any other configuration can, of course, be obtained in a similar way. 
We illustrate these three configurations in figure~\ref{fig:configuration}. For all the results in this paper, we choose the parameters estimated for the Euclid spectroscopic survey, table 3 in~\cite{Euclid:2019clj}. We choose $b_{10}=1.46$ and take the two other biases to be zero ($b_{20}=0$ and $b_{K^2}=0$). As discussed earlier, this does not reduce the generality of our results since all the scale-dependent operators associated with these two bases are already contained in the $\mathcal F_2$ kernel required for the $b_{10}$ result.

For the stochastic terms, we pedagogically choose the leading-order contribution as
\eq{
P_{\eps\eps}(\chi, \chi') = P_{\eps \eps_\delta}(\chi, \chi') = \frac{1}{\bar{n}},
}
where we take the galaxy number density $\bar{n} = 6.86\times 10^{-4}\, h^3\rm{Mpc}^{-3}$~\cite{Euclid:2019clj}.
We do not explicitly include the purely stochastic term from $B_{\epsilon}\approx 1/\bar{n}^2$, since this is scale-independent in 3D, and also when projected to the angular bispectrum. For the equal-time configuration discussed below, the contribution from $B_{\epsilon}$ to the angular bispectrum is around $1\times 10^{-12}$, which is much smaller than other terms on large scales.

\begin{figure}[t]
    \centering    \includegraphics[width=\linewidth]{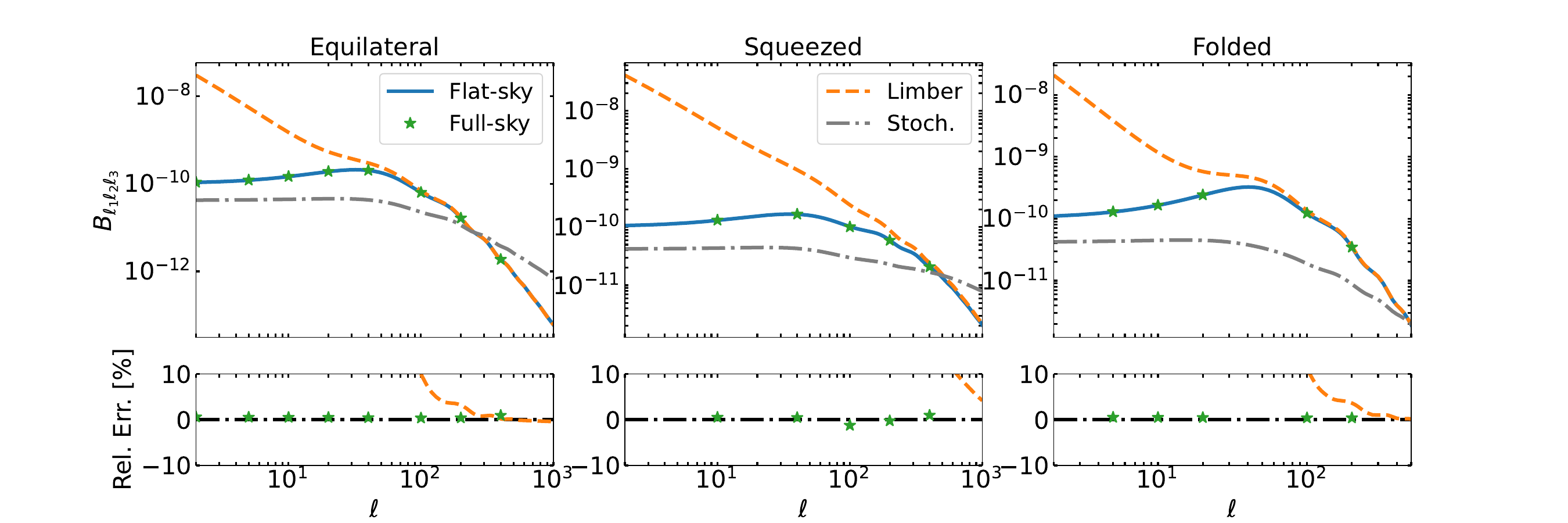}
    \caption{Real space angular bispectrum with \emph{equal-time} radial Gaussian window functions (centered at $z_i=1$ and with $\sigma_z= 0.05$) with $b_{10}=1.46$, $b_{20}=b_{K^2}=0$ and $\bar{n}=6.86\times 10^{-4}\, h^3\rm{Mpc}^{-3}$. From left to right, each panel represents a different configuration: equilateral, squeezed and folded, as illustrated in figure~\ref{fig:configuration}. In all three panels, we compare the flat-sky approximation (blue solid line), the Limber approximation (orange dashed line), and the brute-force full-sky (green stars) results. Note that we do not include the stochastic terms in these lines. Instead, we separately demonstrate the (scale-dependent) stochastic contribution by the dash-dotted gray lines.
    In the bottom panels, we show the fractional residuals compared to the flat-sky results. We note that in the left and the middle panels, we show the multipole range $\ell=2$--$1000$, while for the right panel (folded), we show $\ell=2$--$500$.}
    \label{fig:equal_real}
\end{figure}

In figure~\ref{fig:equal_real}, we show equal-time results for the three bispectrum configurations. We adopt three equal redshift bins centred at $z_1 = z_2 = z_3 = 1.0$, with a narrow Gaussian radial window function ($\sigma_z=0.05$). We find excellent agreement between our flat-sky approximation and the brute-force full-sky computation for all multipoles and configurations. The agreement is better than one percent where the full-sky results are converged and a few percent as we approach higher multipoles, where the full-sky direct integration becomes difficult (and consequently less accurate). We also see that the Limber approximation can only reach this accuracy for multipoles larger than $\ell \approx 250$, with an additional caveat that for our squeezed configuration, one cannot use the canonical Limber approximation (except at very high $\ell$, where the soft mode $1+\ell/10$ becomes large). The partial-Limber approximation developed in appendix~\ref{append:extended_limber} is applicable to lower multipoles; see figure~\ref{fig:extended_limber}.
However, this partial-Limber approximation is computationally less convenient than our flat-sky approximation. The stochastic terms have a non-negligible contribution across all scales. They are almost constant on large scales, while becoming dominant for $\ell\gtrsim 300$. 

\begin{figure}[t]
    \centering
    \includegraphics[width=\linewidth]{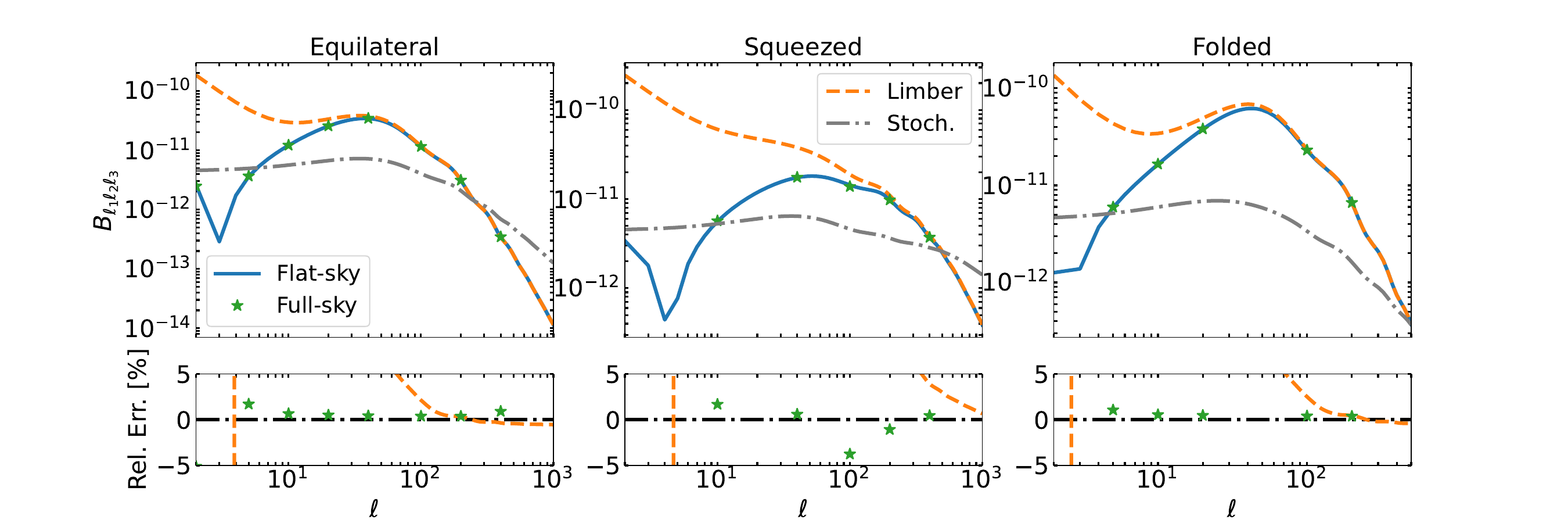}
    \caption{Similar to figure~\ref{fig:equal_real}, but for unequal radial Gaussian window functions with $z_1 = 1.0$, $z_2=1.05$ and $z_3=1.1$ and width $\sigma_z=0.1$. Note that both full-sky and flat-sky angular bispectrum go negative around $\ell=3$ (the absolute value is plotted).
    }
    \label{fig:unequal_real}
\end{figure}

In figure~\ref{fig:unequal_real}, we show results for unequal radial window functions for our three configurations. Here, we set $z_1 = 1.0$, $z_2=1.05$ and $z_3=1.1$, with equal widths of $\sigma_z=0.1$ (broader than adopted for the equal-time case above).
Similarly to the equal-time case, we again see an excellent agreement between our flat-sky approximation and the brute-force full-sky computation for all multipoles and configurations. The agreement is around several percent for $\ell \lesssim 10$ and less than one percent on scales smaller than that. The Limber approximation can achieve such agreement only on scales $\ell \gtrsim 100$ (with the usual caveat for the squeezed configuration). Note, however, that the accuracy of the Limber approximation depends in detail on the central values and widths of the window functions as these determine the degree of overlap. We also notice that the Limber approximation becomes more accurate in this case with broader window function. The broadness of window functions in this unequal-time case is compensating the inaccuracy of Limber approximation due to unequal-time effects.

In general, we see that the canonical Limber approximation converges to the flat-sky results on small scales in all cases, as we would expect. We also note that our full-sky results start to diverge from the flat-sky results on small scales. Given the agreement between the flat-sky and Limber-approximated results on these scales, we expect that inaccuracies in the full-sky computation are becoming apparent there. Since we perform a direct integration over the spherical Bessel functions when calculating the full-sky results, the required numerical accuracy for integrating over the highly oscillatory integrands can be hard to achieve. Thus, we expect the flat-sky numerical results to be more trustworthy and accurate on these small scales than our full-sky computations. However, we note that the small differences at low $\ell$ reflect genuine inaccuracies in the flat-sky approximation.

\section{Angular bispectrum in redshift space}
\label{sec:full_bispec_rsd}

In this section, we add the effects of redshift space distortions (RSD) to the tree-level bispectrum. To do so, we first derive the second-order contributions of RSD to the tracer overdensity field. We then apply these results to compute the RSD contributions to the tree-level angular bispectrum. We extend the flat-sky approximations of the previous section to account for RSD, and compare numerical flat-sky results with full-sky results on large scales (low $\ell$). Finally, we discuss some limitations of the Limber approximation in the presence of RSD.

\subsection{Density fields in redshift space}
\label{subsec:density_fields_RSD}

The redshift space distortions (RSD) are caused by the peculiar velocity of tracers $\vec v$, where the observed redshift is composed of cosmological redshift (due to the Hubble flow) and the kinetic Doppler effect (due to peculiar velocity),
\eq{
1+z_{\rm pec} &= \gamma (1 + v/c) \approx 1 + v/c \, , \\
1+z_{\rm obs} &= (1+z_{\rm cosmo})(1+z_{\rm pec})\, , 
}
where $\gamma$ is the Lorenz factor for $|\vec v|$ and we have temporarily restored factors of the speed of light, $c$. The observed radial position, in a flat $\Lambda$CDM (up to first order in $v/(\mathcal{H}\chi)$, which is an excellent approximation at cosmological distances), is
\eq{
\label{eq:red_mapping}
\vec s = \vec x + \frac{\vec v \cdot \hat{\vec{n}}}{\mathcal{H}} \hat{\vec{n}} = \vec x + f u_{\hat{n}} \hat{\vec{n}} \, ,
}
where $\mathcal{H}$ is the comoving Hubble parameter, $\hat{\vec{n}}$ is the line-of-sight direction, $f \equiv d \ln D / d\ln a$ is the logarithmic growth rate, $\vec{v}$ is the peculiar velocity, and we defined $u_{\hat{n}} \equiv \vec{v} \cdot \hat{\vec{n}}/(f\mathcal{H})$. The projected overdensity field of galaxies within some radial window function\footnote{We assume that selection is in redshift space, i.e., we have a window function $W_z(z_{\text{obs}})$. The equivalent radial window function $W(\chi)$ is defined with the usual unperturbed relation between redshift and comoving distance $\chi$, with the remapping due to RSD taken account of through the redshift space overdensity (see eq.~\ref{eq:jacobian} below).} $W(\chi)$ is
\eq{
n(\hat{\vec{n}}) 
= \int d\chi\,  W(\chi) n_{g,s}(\vec{s};\eta_0-\chi)
= \int_\chi n_{g,s}(\vec{s};\eta_0-\chi) \, ,
}
where $n_{g,s}$ is the observed density in redshift space, and we have taken $\chi$ as the amplitude of $\vec{s}$. Thus, our immediate task is to express the redshift space overdensity $\delta_s$ (defined as $\delta_s \equiv n_s/\bar n_s - 1$, with the mean density $\bar n_s$), in terms of the overdensity in real space $\delta_r$. We can achieve this by imposing the conservation of the number of tracers at some fixed redshift $z$ under the mapping given in eq.~\eqref{eq:red_mapping}. We have
\eq{
\left[1+\delta_{s}(\vec s, z)\right] d^3 \vec{s} = \left[ 1+\delta_r(\vec x, z)\right] d^3 \vec{x}\, ,
\label{eq:jacobian}
}
where we have assumed no evolution of the background density, so that $\bar n_s = \bar n_r$, and ignored other $O(\vec{v}\cdot \hat{\vec{n}}/c)$ terms arising from time derivatives of $f u_{\hat{n}}$ and other kinematic effects; see, e.g.,~\cite{Hamilton:1997} for additional terms that arise from these. 
Expanding eq.~\eqref{eq:jacobian} up to the second order in the field $u_{\hat{n}}$ gives 
\eq{
\label{Eq:delta_s}
1+\delta_s(\vec s) 
&= \int d^3 \vec{x}\,  \left[ 1+\delta_r(\vec x)\right] \delta^{\mathrm{D}}(\vec s - \vec x - f u_{\hat{n}} \hat{\vec{n}}) \non \\
&= \int_{\vec q} \int d^3 \vec{x}\, \left[ 1+\delta_r(\vec x)\right] e^{i\vec q \cdot (\vec s - \vec x - f u_{\hat{n}} \hat{\vec{n}})} \non \\
&= \sum_{n=0}^{\infty} \frac{(f)^n}{n!} \int_{\vec q} e^{i\vec q\cdot \vec s} \int d^3 \vec{x} \left[ 1+\delta_r(\vec x)\right] u_{\hat{n}}^n \partial_x^n e^{-i\vec q \cdot \vec x} \non \\ 
&= \sum_{n=0}^{\infty} \frac{(-f)^n}{n!} \int d^3 \vec{x}\; \frac{1}{x^2}\, \partial_{x}^n \left( x^2\left[ 1+\delta_r(\vec x)\right] u_{\hat{n}}^n\right)\, \delta^{\mathrm{D}}(\vec s - \vec x) \non \\
& \approx \underbrace{1 + \delta_r}_{n=0} \underbrace{-f \lb 1+\delta_r\rb \partial_s u_{\hat{n}} -f (\partial_s \delta_r)\; u_{\hat{n}} - \frac{2}{s}\lb 1+\delta_r\rb f u_{\hat{n}} }_{n=1} \non \\ 
& \hspace{1cm}\underbrace{+f^2\lb 1+\delta_r\rb\left[ (\partial_s u_{\hat{n}} )^2 + u_{\hat{n}} \partial^2_s u_{\hat{n}} \right] + 2f^2(\partial_s \delta_r)\; u_{\hat{n}} \partial_s u_{\hat{n}} + \frac{f^2}{2}(\partial^2_s \delta_r)\; u_{\hat{n}}^2}_{n=2} \non\\
&\hspace{1cm}\underbrace{+\frac{2f^2}{s}\left[ (\partial_s \delta_r)\; u_{\hat{n}}^2 + 2\lb 1+\delta_r\rb u_{\hat{n}} \partial_s u_{\hat{n}}\right] + \frac{f^2}{s^2}\lb 1+\delta_r\rb u_{\hat{n}}^2}_{n=2}\, ,
}
where all the quantities are evaluated at a position $\vec s$, in redshift space. The real-space overdensity can be found in eqs.~(\ref{eq:delta_expansion_in_real_space}--\ref{eq:delta_2_real_space}), while the peculiar velocity term, up to the second order, can be written as
\eq{\label{eq:velocity}
u_{\hat{n}}(\vec x) = - \frac{1}{f \mathcal{H}} \int_{\vec k} \frac{i\vec{k}\cdot \hat{\vec{n}}}{k^2} \lb \theta^{(1)} (\vec k)+\theta^{(2)} (\vec k)\rb e^{i \vec k \cdot \vec x},
}
where $\theta$ is the divergence of the peculiar velocity. 

We consider $\delta_s$ up to the second order in the linear matter density contrast $\delta^{(1)}$, and thus we can organize the redshift space overdensity field perturbatively as $\delta_s = \delta_s^{(1)} + \delta_s^{(2)}$, where
\eeq{
\label{eq:delta_s_1st}
\delta_s^{(1)} = \delta_r^{(1)}  - f \partial_s u_{\hat{n}}^{(1)} - \frac{2f}{s} u_{\hat{n}}^{(1)} \, ,
}
and
\eq{
\label{eq:delta_s_2nd}
\delta_s^{(2)} &= \delta_r^{(2)}  - f \partial_s u_{\hat{n}}^{(2)} - f \delta_r^{(1)}\partial_s u_{\hat{n}}^{(1)} - f \lb\partial_s \delta_r^{(1)}\rb u_{\hat{n}}^{(1)} + f^2 \left(\partial_s u_{\hat{n}}^{(1)} \right)^2 + f^2 u_{\hat{n}}^{(1)} \partial^2_s u_{\hat{n}}^{(1)} \non\\
&\hspace{5cm} - \frac{2f}{s}u_{\hat{n}}^{(2)} - \frac{2f}{s}\delta^{(1)}_r u_{\hat{n}}^{(1)}
+\frac{4f^2}{s} u_{\hat{n}}^{(1)} \partial_s u_{\hat{n}}^{(1)} + \frac{f^2}{s^2} \left(u_{\hat{n}}^{(1)}\right)^2 \, .
}
The term $-2 f u_{\hat{n}}^{(1)}/s$ in eq.~\eqref{eq:delta_s_1st}, as well as all of the terms in the second line of eq.~\eqref{eq:delta_s_2nd}, originate from a variation of the line of sight across the volume, i.e., they come from the expansion of the $x^2/s^2$ terms at the position $\vec s$. Consequently, we can label these terms non-distant-observer corrections (nDOC). In the $\Lambda$CDM universe, as we probe distances $z \gtrsim 0.3$, these observer corrections become negligible (less than around $1\%$ even on the largest scales), and we can rely on the distant-observer approximation. This is equivalent to the flat-sky approximation, i.e., assuming a fixed $\hat{\vec n}$ direction in eq.~\eqref{Eq:delta_s}. In appendix~\ref{append:misalign}, we explicitly show that these terms are indeed negligible in the case when $z = 1.0$, confirming our estimates. Thus, in the rest of this work, we will neglect the nDOC contributions.

We also need to add stochastic contributions to the terms given above. As we saw in the case of real space in section~\ref{sec:fullsky_real}, to the first-order density field we need to add the leading stochastic $\eps^{(1)}$ contribution, while at the second order, besides the purely stochastic term $\eps^{(2)}$, we have a mixed term $\eps^{(1)}_\delta\delta^{(1)}$. In redshift space, there is an additional mixed second-order term $-f \partial_s (\epsilon^{(1)} u^{(1)}_{\hat{n}})$ that also contributes to the tree-level bispectrum.

The velocity field can be related to the linear matter overdensity fields via
\eq{\label{eq:velocity_1_potential}
\theta^{(1)} = -f\mathcal{H}\delta_{\mathrm{G}}\, ,
}
and 
\eq{
\label{eq:velocity_2_potential}
\theta^{(2)} = -f\mathcal{H}\int \frac{d^3 \vec k_1}{(2\pi)^3}\frac{d^3 \vec k_{2}}{(2\pi)^3} \delta^{\rm D}(\vec k  - \vec k_1 - \vec k_2) \mathcal{G}_2(\vec k_1,\vec k_2) \delta_{\mathrm{G}}(\vec k_1;\; z) \delta_{\mathrm{G}}(\vec k_2;\; z)\, ,
}
where the second-order velocity kernel $\mathcal G_2$ is
\eq{
\label{eq:G_kernel}
\mathcal{G}_2(\vec k_1,\vec k_2) &= \frac{3}{7} + \frac{1}{2}\frac{\vec k_1 \cdot \vec k_2}{k_1 k_2} \lb\frac{k_1}{k_2} + \frac{k_2}{k_1}  \rb + \frac{4}{7} \lb \frac{\vec k_1 \cdot \vec k_2}{k_1 k_2}\rb^2 \non \\
& = \frac{3}{14} - \frac{3}{28} \frac{k_1^2}{k_2^2} - \frac{3}{28} \frac{k_2^2}{k_1^2} - \frac{1}{28} \frac{k_3^2}{k_1^2} - \frac{1}{28} \frac{k_3^2}{k_2^2} + \frac{1}{7}\frac{k_3^4}{k_1^2 k_2^2} \, .
} 
Here, in the second line, we give the separable form similar to eqs.~\eqref{eq:F_kernel} and~\eqref{eq:tidal_kernel}.

To compute the angular bispectrum, we need to obtain the harmonic coefficients, as defined in eq.~\eqref{eq:general_decomposition}, for the field in redshift space. Using eqs.~\eqref{eq:delta_s_1st} and~\eqref{eq:delta_s_2nd} we derive the leading-order and next-to-leading-order harmonic coefficients $\delta_{\ell m}^{(1)}$ and $\delta_{\ell m}^{(2)}$ in redshift space. These can be obtained analogously to the real space eqs.~\eqref{eq:delta_1_decomposition_real} and~\eqref{eq:delta_2_decomposition_real}, where the $i\vec{k}\cdot \hat{\vec{n}}$ factors in the velocity can be treated as derivatives of the $\exp(i \chi \vec{k}\cdot \hat{s})$. We thus have
\eq{
\label{eq:delta_s_1st_lm}
\delta_{\ell m}^{(1)} 
&= \int_\chi \int d^2 \hat{\vec{s}}\, Y_{\ell m}^\ast (\hat{\vec{s}}) \int \frac{d^3 \vec{k} }{(2\pi)^3}\, 
\lb  \delta^{(1)}(\vec{k}) - f (i\vec{k}\cdot \hat{\vec n})  u_{\hat{n}}^{(1)}(\vec{k})\rb
e^{i\vec{k}\cdot\vec{s}} \non\\
& = \int_\chi \int d^2 \hat{\vec{s}}\, Y_{\ell m}^\ast(\hat{\vec{s}}) \int \frac{d^3 \vec{k} }{(2\pi)^3}\,  \lb  \delta^{(1)}(\vec{k}) + \theta^{(1)}(\vec{k}) \frac{1}{k^2} \frac{1}{\mathcal{H}} \frac{\partial^2 }{\partial\chi^2}  \rb e^{i\vec{k}\cdot\vec{s}} \non \\
& = 4\pi i^\ell \int_\chi \int \frac{d^3 \vec{k} }{(2\pi)^3}\, Y_{\ell m}^\ast(\hat{\vec{k}}) \lb j_{\ell}\lb k\chi \rb \delta^{(1)}(\vec{k}) + j''_{\ell}\lb k\chi \rb \frac{1}{\mathcal{H}}\theta^{(1)}(\vec k) \rb\, ,
}
where everything is expressed in terms of the real-space linear galaxy overdensity and velocity divergence fields (We drop the subscript $r$ from the real-space density from hereon.). In addition, we have
a single stochastic contribution at first order
\eeq{
\delta_{\ell m, {\rm stoch.}}^{(1)} = 4\pi i^\ell \int_\chi \int \frac{d^3 \vec{k} }{(2\pi)^3}\, Y_{\ell m}^\ast(\hat{\vec{k}}) j_{\ell}\lb k\chi \rb \eps^{(1)}(\vec{k}) \, .
}

Similarly, for the second-order harmonic coefficients, we have 
\eq{
\label{eq:delta_s_2nd_lm}
\delta_{\ell m}^{(2)} 
&= \int_\chi \int d^2 \hat{\vec{s}}\, Y_{\ell m}^\ast(\hat{\vec{s}}) \Bigg[ \int \frac{d^3 \vec{k} }{(2\pi)^3}\, \lb  \delta^{(2)}(\vec{k})  - f (i\vec{k}\cdot \hat{\vec{n}}) u_{\hat{n}}^{(2)}(\vec{k}) \rb 
e^{i\vec{k}\cdot\vec{s}} \non \\
&\hspace{1cm} + 
\int \frac{d^3 \vec{k}_1 }{(2\pi)^3} \frac{d^3 \vec{k}_2 }{(2\pi)^3}\,
\bigg( - f \left[ i \lb \vec{k}_1  + \vec{k}_2 \rb \cdot \hat{\vec{n}} \right]  \delta^{(1)}(\vec k_1) u_{\hat{n}}^{(1)}(\vec k_2) \non \\
& \hspace{3cm} + f^2 \left[ (i\vec{k}_1\cdot \hat{\vec{n}})(i\vec{k}_2\cdot \hat{n}) + (i\vec{k}_2\cdot \hat{\vec{n}})^2 \right] u_{\hat{n}}^{(1)} (\vec k_1) u_{\hat{n}}^{(1)}(\vec k_2) \bigg) e^{i \lb \vec k_1 + \vec k_2 \rb \cdot \vec s} \Bigg] \, .
}
We can rewrite the first line above in the same way as at linear order, i.e., 
\eeq{
\lb  \delta^{(2)}(\vec{k})  - f (i\vec{k}\cdot \hat{\vec n}) u_{\hat{n}}^{(2)}(\vec{k}) \rb e^{i\vec{k}\cdot\vec{s}}
= \lb \delta^{(2)}(\vec{k})  + \theta^{(2)}(\vec{k}) \frac{1}{k^2}\frac{1}{\mathcal{H}}\frac{\partial^2 }{\partial\chi^2}  \rb e^{i\chi \vec{k}\cdot\vec{\hat n}} \, ,
}
while the remaining part, in the second and third lines, can be written as 
\eq{
&\bigg( - f \lb i \vec{k}_{12} \cdot \hat{\vec n} \rb  \delta^{(1)}(\vec k_1) u_{\hat{\vec n}}^{(1)}(\vec k_2) - f^2 \left[ (\vec{k}_1\cdot \hat{\vec n})(\vec{k}_2\cdot \hat{\vec n}) + (\vec{k}_2\cdot \hat{\vec n})^2 \right] u_{\hat{n}}^{(1)} (\vec k_1) u_{\hat{n}}^{(1)}(\vec k_2) \bigg) e^{i \vec k_{12} \cdot \vec s} \non\\
&\hspace{1cm} = \Bigg[ \frac{1}{\mathcal{H}} \frac{1}{k_2^2} \lb \frac{\partial^2 }{\partial \chi_2^2} + \frac{\partial^2 }{\partial \chi_1 \partial \chi_2} \rb  
e^{i \lb \chi_1 \vec k_1 + \chi_2 \vec k_2 \rb \cdot \hat{\vec n}}
\delta^{(1)}(\vec k_1) \theta^{(1)}(\vec k_2) \non \\
&\hspace{2.2cm} + \frac{1}{\mathcal{H}^2}  \frac{1}{k_1^2 k_2^2} \lb \frac{\partial^4}{\partial \chi_1^2 \partial \chi_2^2} + \frac{\partial^4 }{\partial \chi_1 \partial \chi_2^3} \rb
e^{i \lb \chi_1 \vec k_1 + \chi_2 \vec k_2 \rb \cdot \hat{\vec n}}
 \theta^{(1)}(\vec k_1) \theta^{(1)}(\vec k_2) \Bigg]_{\chi_1 = \chi_2 = \chi} \, ,
}
where $\vec{k}_{12}\equiv \vec{k}_1+\vec{k}_2$ and we used the fact that $\vec s = \chi \hat{\vec n}$.
Combining all these, and using the plane-wave expansion, we obtain
\eq{
\label{eq:delta_s_2nd_lm_II}
\delta_{\ell m}^{(2)} 
&= 4\pi i^{\ell} \int_\chi \int \frac{d^3 \vec{k} }{(2\pi)^3} Y_{\ell m}^\ast(\hat{\vec{k}}) \bigg(  \underbrace{j_{\ell}(k\chi) \delta^{(2)}(\vec{k})}_{\rm SOD}  + \underbrace{ j''_{\ell}(k\chi) \frac{1}{\mathcal{H}} \theta^{(2)}(\vec{k})}_{\rm SOV} \bigg) \non \\
&\hspace{0.2cm} + (4\pi)^2 \sum_{\ell_1 m_1} \sum_{\ell_2 m_2}i^{\ell_1+\ell_2} \mathcal{G}_{\ell \ell_1 \ell_2}^{m m_1 m_2}
\int_{\chi} \int \frac{d^3 \vec{k}_1}{(2\pi)^3} \frac{d^3 \vec{k}_2}{(2\pi)^3}\,  Y_{\ell_1 m_1}^\ast(\hat{\vec k}_1) Y_{\ell_2 m_2}^\ast(\hat{\vec k}_2) \non \\
&\hspace{0.2cm} \times \frac{1}{2}\lb  \underbrace{\frac{1}{\mathcal{H}}
\lb j_{\ell_1}(k_1 \chi) j''_{\ell_2}(k_2 \chi) + k_1 j'_{\ell_1}(k_1 \chi) \frac{1}{k_2} j'_{\ell_2}(k_2 \chi) \rb \delta^{(1)}(\vec k_1)\theta^{(1)}(\vec k_2) + \mathrm{perm}_{1\leftrightarrow 2}}_{\rm DV} \right.  \non \\
&\hspace{0.6cm} \left. + \underbrace{\frac{1}{\mathcal{H}^2} \lb j''_{\ell_1}(k_1 \chi) j''_{\ell_2}(k_2 \chi) + \frac{1}{k_1}j'_{\ell_1}(k_1 \chi) k_2 j'''_{\ell_2}(k_2 \chi) \rb \theta^{(1)} (\vec k_1) \theta^{(1)}(\vec k_2) + \mathrm{perm}_{1\leftrightarrow 2} }_{\rm VV}\rb \, ,
}

where in the last two lines we have symmetrised in $\vec{k}_1$ and $\vec{k}_2$ for later convenience.
We can identify four different components in the second-order redshift-space tracer overdensity: second-order density (SOD), second-order velocity (SOV), density--velocity (DV) and velocity--velocity (VV). It is convenient to write down the bispectrum separately with these four terms. As is evident from these results, the velocity terms introduce derivatives of the spherical Bessel functions. In fact, the Bessel function terms in the SOV, DV and VV components are all total derivatives with respect to $\chi$, up to time derivatives of the real-space density and velocity divergence fields. This behaviour is inherited from the relevant terms in the first line of eq.~\eqref{eq:delta_s_2nd}. As at first order, this suppresses the RSD contributions to the second-order projected density at high $\ell$ after integrating over broad window functions (i.e., the Limber limit). 

Stochastic contributions at second order give
{\allowdisplaybreaks
\eq{
\delta_{\ell m, {\rm stoch.}}^{(2)} 
&= 4\pi i^{\ell} \int_\chi \int \frac{d^3 \vec{k}}{(2\pi)^3} Y_{\ell m}^\ast(\hat{\vec{k}}) j_{\ell}(k\chi) \eps^{(2)}(\vec{k}) \non \\
&\hspace{0cm} + (4\pi)^2 \sum_{\ell_1 m_1} \sum_{\ell_2 m_2}i^{\ell_1+\ell_2} \mathcal{G}_{\ell \ell_1 \ell_2}^{m m_1 m_2}
\int_{\chi} \int \frac{d^3 \vec{k}_1}{(2\pi)^3} \frac{d^3 \vec{k}_2}{(2\pi)^3}\,  Y_{\ell_1 m_1}^\ast(\hat{\vec{k}}_1) Y_{\ell_2 m_2}^\ast(\hat{\vec{k}}_2) \non \\
&\hspace{0cm} \times \frac{1}{2} \Bigg[ j_{\ell_1}(k_1 \chi) j_{\ell_2}(k_2 \chi) \eps_{\delta}^{(1)}(\vec k_1)\delta^{(1)}(\vec k_2) + \mathrm{perm}_{1\leftrightarrow 2}  \non \\
&\hspace{0cm} + \frac{1}{\mathcal{H}}
\lb j_{\ell_1}(k_1 \chi) j''_{\ell_2}(k_2 \chi) + k_1 j'_{\ell_1}(k_1 \chi) \frac{1}{k_2} j'_{\ell_2}(k_2 \chi) \rb \eps^{(1)}(\vec k_1)\theta^{(1)}(\vec k_2) + \mathrm{perm}_{1\leftrightarrow 2}\Bigg] \, .
}
}

\subsection{Angular bispectrum in redshift space}

Given the results for the first- and second-order harmonic coefficients of the redshift space overdensity, we proceed to obtain the angular bispectrum in redshift space. Similarly to the derivations presented in eq.~\eqref{eq:B_real} for real space, we can identify four contributions to the redshift space bispectrum arising from the different terms contributing to the second-order eq.~\eqref{eq:delta_s_2nd_lm_II}. We can write 
\eeq{
B_{\ell_1\ell_2\ell_3}
= B_{\ell_1\ell_2\ell_3}^{\rm SOD} 
+ B_{\ell_1\ell_2\ell_3}^{\rm SOV} 
+ B_{\ell_1\ell_2\ell_3}^{\rm DV} 
+ B_{\ell_1\ell_2\ell_3}^{\rm VV}
+ B_{\ell_1\ell_2\ell_3}^{\rm Stoch.}\, ,
\label{eq:RSDbreakdown}
}
where each of the components is given explicitly below.

1. \emph{Second-order density}
\eq{
\label{eq:3pt_correlator_SOD}
B_{\ell_1\ell_2\ell_3}^{\rm SOD} 
&= 2 \lb \frac{2}{\pi}\rb^3 \int_{\chi_1\chi_2\chi_3} D_1 D_2 D_3^2
\int r^2 dr \int k_1^2 dk_1 k_2^2 dk_2 k_3^2 dk_3\; \mathcal{K}_2(\vec{k}_1, \vec{k}_2;\; \chi_3)  P_L(k_1)P_L(k_2) \non \\
&\hspace{3.2cm} \times \lb b_{10}j_{\ell_1}(k_1\chi_1) - f_1 j''_{\ell_1}(k_1\chi_1) \rb \lb b_{10}j_{\ell_2}(k_2\chi_2) - f_2 j''_{\ell_2}(k_2\chi_2)\rb j_{\ell_3}(k_3 \chi_3) \non\\
&\hspace{3.2cm}\times j_{\ell_1}(k_1r) j_{\ell_2}(k_2r) j_{\ell_3}(k_3r) + \text{2 perms.}
}

2. \emph{Second-order velocity}
\eq{\label{eq:3pt_correlator_SOV}
B_{\ell_1\ell_2\ell_3}^{\rm SOV} &= -2\lb \frac{2}{\pi}\rb^3 \int_{\chi_1\chi_2\chi_3} D_1 D_2 D_3^2 f_3\int r^2 dr \int k_1^2 dk_1 k_2^2 dk_2 k_3^2 dk_3\; \mathcal{G}_2(\vec{k}_1, \vec{k}_2) P_L(k_1)P_L(k_2) \non \\
&\hspace{3.0cm} \times
\lb b_{10}j_{\ell_1}(k_1\chi_1) - f_1 j''_{\ell_1}(k_1\chi_1) \rb \lb b_{10}j_{\ell_2}(k_2\chi_2) - f_2 j''_{\ell_2}(k_2\chi_2) \rb j''_{\ell_3}(k_3 \chi_3)\non\\
&\hspace{3.0cm}\times j_{\ell_1}(k_1r) j_{\ell_2}(k_2r) j_{\ell_3}(k_3r) + \text{2 perms.}
}

3. \emph{Density-velocity}
\eq{\label{eq:3pt_correlator_DV}
B_{\ell_1\ell_2\ell_3}^{\rm DV} &= -\lb \frac{2}{\pi}\rb^2 \int_{\chi_1\chi_2\chi_3} D_1 D_2 D_3^2 f_3 \int k_1^2 dk_1 k_2^2 dk_2\; P_L(k_1)P_L(k_2) \non\\
& \hspace{1.5cm} \times \lb j_{\ell_1}(k_1\chi_3) j''_{\ell_2}(k_2\chi_3) 
+ k_1 j'_{\ell_1}(k_1 \chi_3) \frac{1}{k_2} j'_{\ell_2}(k_2 \chi_3) 
+ \mathrm{perm}_{1\leftrightarrow 2}
\rb \non \\
& \hspace{1.5cm} \times \lb b_{10}j_{\ell_1}(k_1\chi_1) - f_1 j''_{\ell_1}(k_1\chi_1) \rb \lb b_{10}j_{\ell_2}(k_2\chi_2) - f_2 j''_{\ell_2}(k_2\chi_2)\rb + \text{2 perms.}
}

4. \emph{Velocity-velocity}
\eq{\label{eq:3pt_correlator_VV}
B_{\ell_1\ell_2\ell_3}^{\rm VV} &= \lb \frac{2}{\pi}\rb^2 \int_{\chi_1\chi_2\chi_3} D_1 D_2 D_3^2 f^2_3 \int k_1^2 dk_1 k_2^2 dk_2\; P_L(k_1)P_L(k_2) \; \non\\
& \hspace{1cm} \times\lb k_1 j'''_{\ell_1}(k_1\chi_3) \frac{1}{k_2}j'_{\ell_2}(k_2\chi_3) + \frac{1}{k_1}j'_{\ell_1}(k_1\chi_3) k_2 j'''_{\ell_2}(k_2\chi_3) + 2j''_{\ell_1}(k_1 \chi_3) j''_{\ell_2}(k_2 \chi_3) \rb \non \\
& \hspace{1cm} \times \lb b_{10}j_{\ell_1}(k_1\chi_1) - f_1 j''_{\ell_1}(k_1\chi_1) \rb \lb b_{10}j_{\ell_2}(k_2\chi_2) - f_2 j''_{\ell_2}(k_2\chi_2) \rb + \text{2 perms.}
}

5. \emph{Stochastic}
\eq{
B_{\ell_1\ell_2\ell_3}^{\rm Stoch.} 
&= \lb \frac{2}{\pi} \rb^2 
\int_{\chi_1,\chi_2,\chi_3} D_1 D_3
 \int k_1^2 d k_1 k_2^2 d k_2\, P_L(k_1)  \non\\
& \hspace{1cm} \times  \Big( b_{10} (\chi_1) j_{\ell_1}\lb k_1\chi_1 \rb - f_1 j''_{\ell_1}\lb k_1\chi_1 \rb  \Big)
j_{\ell_2}\lb k_2\chi_2 \rb \non \\
& \hspace{1cm} \times\bigg(j_{\ell_1}(k_1 \chi_3) j_{\ell_2}(k_2 \chi_3) P_{\eps \eps_\delta}(\chi_2,\chi_3) \non\\
& \hspace{1.0cm} - f_3 \lb j_{\ell_2}(k_2 \chi_3) j''_{\ell_1}(k_1 \chi_3) + k_2 j'_{\ell_2}(k_2 \chi_3) \frac{1}{k_1} j'_{\ell_1}(k_1 \chi_3)\rb P_{\eps\eps}(\chi_2,\chi_3) \bigg) + (1\leftrightarrow2) \non\\
& \hspace{1cm} + \text{2 perms.}
} 
We note, again, that we do not explicitly show the scale-invariant stochastic contribution to the angular bispectrum from $B_\epsilon$, which is a constant, and is not affected by the RSD effect.
The SOD term contains the real space angular bispectrum (discussed in section~\ref{sec:fullsky_real}), with additional (Kaiser~\cite{Kaiser1987}) terms in the first-order density field when considering the RSD effect. The SOV term contains the second-order velocity contributions, while the DV and VV terms take care of the second-order density--velocity and velocity--velocity contributions.

All these tree-level terms can be written in a separable form similar to eq.~\eqref{eq:Blll_nnn_real}, the only difference being the partial derivatives acting on the spherical Bessel functions. We can treat these by taking the derivatives of the unprojected angular power spectrum
\eq{
\frac{2}{\pi} \int k^2 dk\; j_{\ell}(k r) j^{(m)}_{\ell}(k\chi) k^{2n} P_L(k) &= \frac{\partial^m}{\partial \chi^m}\mathbb{C}_{n-\frac{m}{2}}(\ell, r, \chi)\, .
}
This treatment allows us again to use our flat-sky approximation described in section~\ref{subsec:numerical_evaluation}. The derivatives can then be treated analytically, acting directly on the special functions in the expansion given in eqs.~\eqref{eq:Expansion} and \eqref{eq:M0i}. Alternatively, they can be taken numerically by first evaluating $\mathbb{C}_{n-\frac{m}{2}}$. We adopt the latter approach in this work.  

Furthermore, the SOD and SOV terms involve an integral over the $k_3$ momentum while lacking a dependence on the linear power spectrum. This is expected, as similar terms also appear in the real space case, giving rise to $\mathbb{I}_n$ terms as defined in eq.~\eqref{eq:unprojected_Cn_real}. The origin of these contributions to the SOD and SOV terms lies in the non-trivial structure of the $\mathcal{K}$ and $\mathcal{G}$ kernels. When expressed in a separable form, these kernels introduce an additional dependence on $k_3$, leading to $\mathbb{I}_n$ contributions for $n \neq 0$. Notably, the $\mathbb{I}_n$ terms are absent in the DV and VV contributions, as only $\mathbb{I}_0$ contributes in these cases. This integral is trivial and can be evaluated directly, thus simplifying these terms.

Moreover, the SOV term contains an additional second-order derivative of the spherical Bessel function, i.e., we have (generalizing to higher derivatives)
\eq{
\frac{2}{\pi} \int k^2 dk\; j_{\ell}(k r) j^{(m)}_{\ell}(k\chi) k^{2n} = \frac{\partial^m}{\partial \chi^m}\mathbb{I}_{n-\frac{m}{2}}(\ell, r, \chi)\, .
}
The $k^{2n}$ factor in the integrand arises from the $\mathcal{G}_2$ kernel; the index $n$ can take values of $0,1,2$. However, the action of the second-order derivative effectively reduces the index of $\mathbb{I}_{n-\frac{m}{2}}$ below zero for $n=0$, which requires a separate treatment. We instead use
\eq{
\label{eq:dd_I_n-1}
\frac{\partial^2}{\partial \chi^2}\mathbb{I}_{-1}(\ell, r, \chi) &= \lb \frac{2}{\pi} \rb \frac{\partial^2}{\partial \chi^2} \int dk\; j_{\ell}(k r) j_{\ell}(k\chi) \non \\
&=  \frac{1}{2\ell+1} \frac{\partial^2}{\partial\chi^2}\lb \chi^{-\ell-1} r^{\ell} \Theta_{\mathrm{H}}(\chi-r) + r^{-\ell-1} \chi^{\ell} \Theta_{\mathrm{H}}(r-\chi) \rb\, ,
}
where $\Theta_{\mathrm{H}}(\chi-r)$ is the Heaviside step function, with $\Theta_{\mathrm{H}}(0)=1/2$. For the remaining cases we have
\eq{
\label{eq:dd_I_n0}
\frac{\partial^2}{\partial \chi^2}\mathbb{I}_{0}(\ell, r, \chi) &= \lb \frac{2}{\pi} \rb \frac{\partial^2}{\partial \chi^2} \int k^2 dk\; j_{\ell}(k r) j_{\ell}(k\chi) = \frac{\partial^2}{\partial \chi^2} \lb \frac{1}{r\chi}\delta^{\rm D}(\chi - r) \rb \, , \\
\label{eq:dd_I_n1}
\frac{\partial^2}{\partial \chi^2}\mathbb{I}_{1}(\ell, r, \chi) &= \lb \frac{2}{\pi} \rb \frac{\partial^2}{\partial \chi^2} \int k^2 dk\; k^2j_{\ell}(k r) j_{\ell}(k\chi) \non \\
&= \frac{\partial^2}{\partial \chi^2} \left[ -\frac{\partial^2}{\partial \chi^2} -\frac{2}{\chi}\frac{\partial}{\partial \chi}+\frac{\ell(\ell+1)}{\chi^2} \right] \lb \frac{1}{\chi^2}\delta^{\rm D}(\chi - r) \rb.
}
In this work, we choose to use integration-by-parts to move the $\chi$ derivatives onto the window function when dealing with eqs.~(\ref{eq:dd_I_n-1}--\ref{eq:dd_I_n1}). This is effective for the Gaussian radial window functions, which we have chosen for an ideal galaxy-clustering scenario. However, we emphasize again that for a non-smooth window function, or a window function that does not go to zero on the boundaries, one needs to be careful with the boundary terms or, alternatively, use integration-by-parts to transfer the derivatives to the unequal-time angular power spectra, as discussed in section~\ref{subsec:numerical_evaluation}.

\subsection{Limber approximation in redshift space}
\label{subsec:Limber_rsd}

We can ignore all terms involving derivatives of the spherical Bessel functions in the Limber limit, as explained after eq.~\eqref{eq:delta_s_2nd_lm_II}. In other words, we can ignore redshift space distortions in this limit and simply use the Limber approximation in real space (eq.~\ref{eq:bispec_real_explicit_limber}). We verify this numerically in section~\ref{subsec:results_rsd}. Partial-Limber approximations for squeezed configurations can be found in appendix~\ref{append:extended_limber}.

\subsection{Results}
\label{subsec:results_rsd}

\begin{figure}
    \centering
    \includegraphics[width=\linewidth]{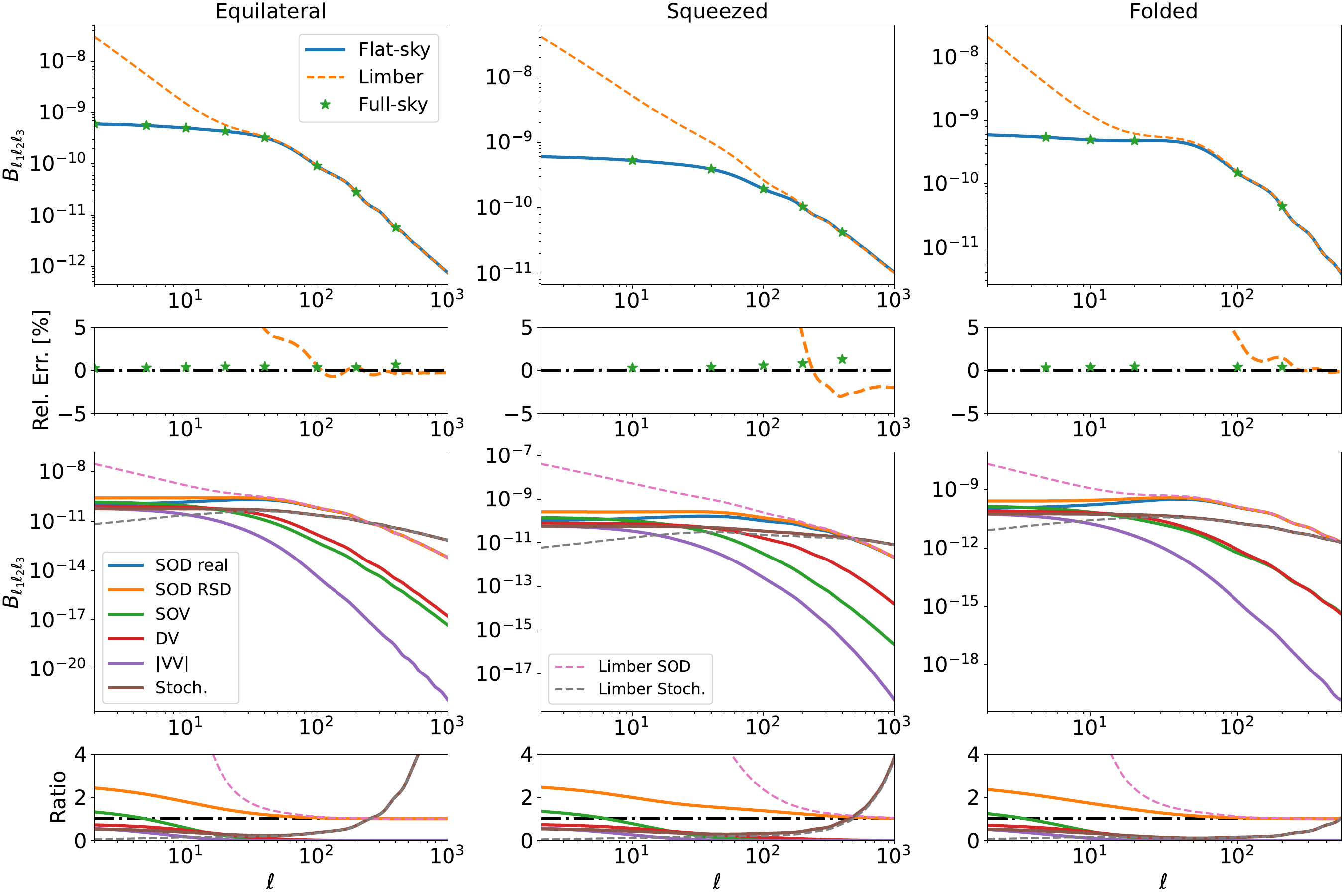}
    \caption{Angular bispectrum including the RSD effect for \textit{equal-time} radial Gaussian window functions centred on $z_1=z_2=z_3=1.0$ and with equal widths $\sigma_z = 0.05$. These, and the bias and mean galaxy number density, are the same as in figure~\ref{fig:equal_real}. The columns show the equilateral, squeezed and folded configurations (left to right). In the first row, we show the tree-level angular bispectrum, where we compare the flat-sky (blue solid line), the Limber approximation (orange dashed line) and the full-sky (green stars) results. (We note that these results include the scale-dependent stochastic contributions.) In the second row, we show the fractional differences relative to the flat-sky results, showing the excellent agreement between the flat- and full-sky results. In the third row, the solid lines show the contributions to the flat-sky results from the different redshift components in eq.~\eqref{eq:RSDbreakdown}. For the SOD term, we show the full contribution (SOD RSD; orange) as well as the part without RSD (SOD real; blue), which is the same as in 
    figure~\ref{fig:equal_real}. The SOV (green), DV (red), magnitude of VV (purple)
    and stochastic (Stoch.; brown) terms are also shown. We also show the Limber approximations to the SOD term (pink dashed) and to the stohastic term (Limber Stoch.; gray dashed). The final row shows the ratio of all these components to SOD real to demonstrate the scale dependence of the different RSD terms.}
    \label{fig:equal_time_rsd}
\end{figure}

\begin{figure}
    \centering
    \includegraphics[width=\linewidth]{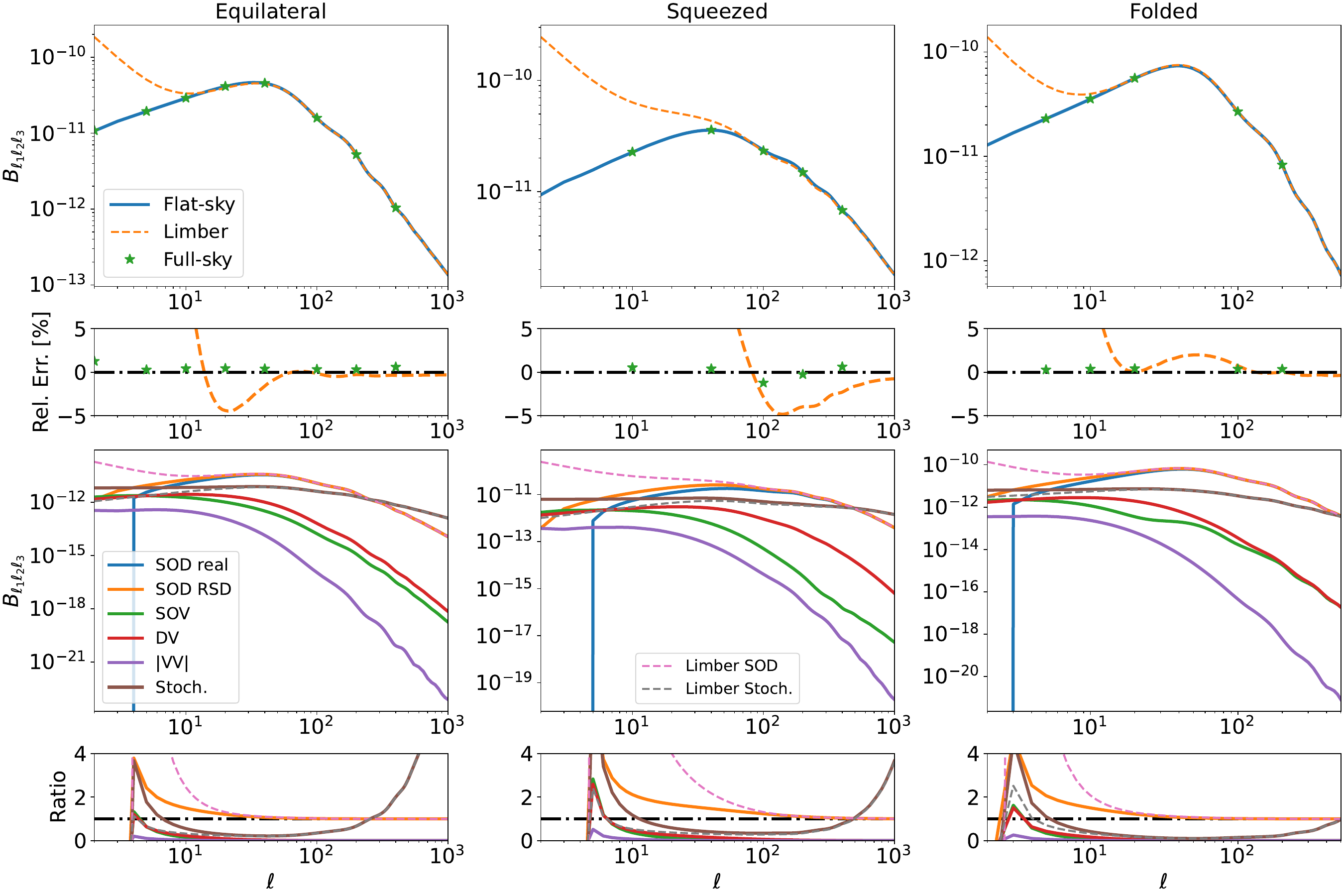}
    \caption{Angular bispectrum with RSD included for \textit{unequal-time} radial Gaussian window functions centred on $z_1=1.0$, $z_2=1.05$ and $z_3=1.1$ with equal widths $\sigma_z = 0.1$. The description of the lines and panels is the same as for figure~\ref{fig:equal_time_rsd}. Note, however, that we show here the absolute value of the VV term since it is negative across all scales.}
    \label{fig:unequal_time_rsd}
\end{figure}

We consider the same three bispectrum configurations that were discussed in section~\ref{subsec:results_realspace}, and we again look at the equivalent equal- and unequal-time setups. We show the equal-time results in figure~\ref{fig:equal_time_rsd}, while the unequal-time results are shown in figure~\ref{fig:unequal_time_rsd}. In all cases, we compare the full-sky results (obtained by brute-force numerical integration for low multipoles) with our flat-sky approximation, finding percent-level agreement over all scales of interest. The exceptions are scales $\ell \gtrsim 200$ in the unequal-time squeezed configuration, where the agreement is several percent, which is consistent with the expected errors of our brute-force numerical integration for these multipoles. These findings are consistent with those obtained for the real space bispectrum, presented in section~\ref{subsec:results_realspace}. Moreover, on small scales, the Limber approximation converges to the flat-sky results, confirming the accuracy of these results. Percent-level agreement of the two is typically achieved around $\ell \gtrsim 100$ in all of our configurations.

As already discussed in section~\ref{subsec:results_realspace}, the unequal-time angular bispectrum without RSD (the solid blue lines in the third row of figure~\ref{fig:unequal_time_rsd}, labelled SOD real) crosses zero on very large scales.
We see that the full SOD term including RSD (solid orange lines labelled SOD RSD in the figure), along with the SOV and DV terms, are all positive, however, while the VV term is negative on all scales (the magnitude is shown in the figure). Nonetheless, the total RSD angular bispectrum remains positive. In both the equilateral and folded configurations, and for the equal-time and unequal-time cases, the SOD RSD converges to SOD real on small scales. This is expected as the RSD corrections are suppressed on scales small compared to the radial width of the window functions; from figures~\ref{fig:equal_time_rsd} and~\ref{fig:unequal_time_rsd}, we see that the linear Kaiser corrections (the terms involving $j''_{\ell}(k\chi)$ in eq.~\ref{eq:3pt_correlator_SOD}) start to become negligible around $\ell\sim 100$. For $z=1$, this multipole corresponds to a scale about $\ell/\chi = k\approx 0.04\, h\mathrm{Mpc}^{-1}$, related to the width $\sigma_z\approx 0.05$, which is consistent with the numerical results.
Similarly, the other RSD-generated terms, SOV, DV and VV, become negligible on small scales.
On the other hand, on large scales the RSD corrections can be significant, with 
The VV term, however, is a small contributor on almost all scales (except for the largest scales in the equal-time case). Even though these trends are, in detail, dependent on the specific window function configurations, they provide a qualitative insight into the relative importance of the various terms. 

Lastly, let us look at the squeezed-limit results. Here, RSD contributions to the SOD term in the bispectrum remain important to larger multipoles than in the other bispectrum configurations. (The SOD RSD in orange in the lower middle panels in figures~\ref{fig:equal_time_rsd} and~\ref{fig:unequal_time_rsd} is above the SOD real, in blue, on all scales.)
The reason for this is that there is always a permutation where the linear RSD Kaiser term is on the soft mode and this is less suppressed by integration through the radial window function.
The SOV term behaves in the squeezed limit similarly to the other configurations: it is important on large scales but drops off rapidly on smaller scales. This is because an appreciable small-scale SOV bispectrum in the squeezed configuration would require the second-order velocity to be associated with one of the two hard modes, but as this is a total radial derivative it is then suppressed by the window-function integration.
The DV term has similar large-scale behaviour, reaching the same amplitude as SOD real for $\ell\lesssim 10$. However, on smaller scales, the DV term is sub-dominant but not negligible. 

\section{Conclusion}
\label{sec:conclusion}

Computation of the full-sky angular bispectrum is a difficult task due to its inherent complexity and the lack of efficient, robust implementations. In Ref.~\cite{Gao_2023}, we demonstrated that the flat-sky approximation of the angular power spectrum can be rigorously derived from the full-sky formalism as the leading-order asymptotic expansion under the distant-observer limit and small survey angles. Building on these results, we showed how the FFTLog expansion \cite{Hamilton:1999} can be utilized to facilitate the fast and accurate computation of the unequal-time angular power spectrum $C_\ell$ (see \cite{Gao_2024}). Here, we extended these methodologies to compute the angular bispectrum within the flat-sky approximation, including the effect of redshift space distortions, 
using tree-level perturbative results for general biased tracers. We also carefully discuss how the stochasticity enters the angular bispectrum and calculated its contribution.

In the first part of this work, we started by neglecting redshift space distortions for simplicity of presentation. We rederived the tree-level full-sky angular bispectrum (analogous to the results presented in \cite{Assassi2017}), which requires the second-order perturbative expansion of the biased tracer field. The angular bispectrum was decomposed into the survey-independent unequal-time bispectrum, \(\mathbb{B}_{\ell_1 \ell_2 \ell_3}\) (the unprojected angular bispectrum), and the integration over arbitrary survey-dependent window functions. As also shown in \cite{Assassi2017}, the separability of the tree-level bispectrum stems from the properties of the second-order biasing kernel, $\mathcal{K}_2$. This allows the angular bispectrum to be expressed as a sum of products of two angular power spectrum terms, $\mathbb{C}_{n}(\ell, r, \chi)$, each involving an integral over wavenumber of a modified linear power spectrum multiplied by a given $k^n$ factor, along with an additional radial integration over one of the distance variables in $\mathbb{C}_{n}(\ell, r, \chi)$. To ensure computational efficiency and robust numerical evaluation, we approximate these constituent angular power spectra, $\mathbb{C}_n$, using the flat-sky approximation described in \cite{Gao_2024}. This step is critical to achieving an accurate and efficient evaluation of the angular bispectrum. In presenting our results, we examined both equal-time and unequal-time (at and near to $z=1$) radial window function scenarios while analyzing three characteristic bispectrum configurations: equilateral, squeezed, and folded. Across all cases, we observed excellent agreement, at the percent level, between brute-force full-sky calculations and our flat-sky approximation. Furthermore, in the small-scale regime, where direct numerical evaluation of the full-sky results becomes infeasible, we demonstrated the convergence of our flat-sky results with the Limber approximation. This reaffirms the accuracy and practical utility of our approach for efficient bispectrum computations on all scales (at the redshifts we considered).

In the second part of this paper, we extended our analysis to include redshift space distortions. Starting with the full-sky bispectrum, we derived the tree-level results by expanding the tracer field to second order. Under the distant-observer approximation, we classifed the angular bispectrum contributions into four components: \emph{second-order density}, \emph{second-order velocity}, \emph{density--velocity}, and \emph{velocity--velocity} terms. The latter three arise only because of redshift space distortions.
Similarly to the case without redshift space distortions, each component can be expressed in a separable form. Redshift-space distortions introduce additional correlations in the angular bispectrum, which can also be represented as products of two $\mathbb{C}_n(\ell)$ terms, now also modified by several radial derivatives. Using our flat-sky approximation for these terms, we computed results for both equal-time and unequal-time radial scenarios (near $z=1$) across equilateral, squeezed, and folded bispectrum configurations. Once again, we found excellent agreement, at the percent level, between the full-sky and flat-sky results across nearly all scales of interest. As when neglecting redshift space distortions, the brute-force computation of full-sky results becomes unreliable at small scales, whereas our flat-sky results were shown to converge with the Limber approximation. This convergence establishes the accuracy and robustness of our flat-sky results across all relevant scales (at the considered redshift), demonstrating their utility for efficient and precise bispectrum analysis.

In examining the limitations of the Limber approximation for the angular bispectrum, we first observed that its reliability on small scales largely arises from the dominance of the second-order density (SOD) term over other contributions, which are suppressed by integrating over radial window functions, in most bispectrum configurations. Additionally, redshift space distortion effects become important
only on relatively larger scales, precisely where the Limber approximation begins to fail. In the squeezed configuration, when one mode is much softer than the others, the canonical Limber approximation is inherently inaccurate (except at high multipoles where our choice of soft leg, $1+\ell/10$, becomes sufficiently large). We present an extension of the Limber approximation for the squeezed configuration in appendix~\ref{append:extended_limber}, which achieves better performance but is more complicated to evaluate.
Beyond these observations, which are primarily based on tree-level results, we anticipate that on sufficiently small scales nonlinear contributions will become significant. 

To the best of our knowledge, the methods presented in this work provide the most efficient approach for computing the galaxy angular bispectrum across a wide range of scales and redshifts of interest. Regarding computational performance, evaluating the $\mathbb{C}_{n}(\ell, r, \chi)$ at $100$ points in $r$ and $200$ in $\chi$,
and encompassing all three permutations, our flat-sky approach (utilizing 400 FFTLog frequencies, which is a very conservative choice; reducing this number would significantly reduce the computational time) requires only 30 seconds per triplet on a single CPU thread ($2.3\,\text{GHz}$). In contrast, the direct brute-force numerical evaluation of the full-sky results takes tens of hours. Specifically, achieving percent-level accuracy with the full-sky approach required 17 hours for low multipoles and up to 29 hours for high multipoles, even when parallelized across three CPU threads. To accompany this paper, we are releasing our Python implementation designed for the tree-level angular bispectrum of galaxy clustering including redshift space distortions\footnote{https://github.com/GZCPhysics/AngularBispectrum\_RSD.git}. This implementation facilitates efficient and optimized usage of the angular bispectrum in MCMC analyses for upcoming spectroscopic and photometric surveys. Additionally, we believe the computational performance of our approach could be further improved through enhanced coding practices, with substantial algorithmic gains possible by adopting optimized decompositions for the linear power spectrum, as demonstrated in implementations such as COBRA \cite{Bakx:2024}. Since \cite{Bakx:2024} has shown that using about $14$ eigenfunctions is enough for representing the 3D matter power spectrum, we estimate an improvement in computational time by around a factor of 14 with COBRA implemented.

\begin{acknowledgments}
Z.G. acknowledges support from the \emph{Agence Nationale de la Recherche} (ANR) under grant No. ANR-23-CPJ1-0160-01. Z.V. acknowledges the support of the Kavli Foundation and Croatian Science Foundation (grant number IP-2025-02-1338). A.C.\ acknowledges support from the STFC (grant number ST/S000623/1).
\end{acknowledgments}  

\appendix

\section{Non-distant-observer correction terms}
\label{append:misalign}

In the main text, we ignored the non-distant-observer corrections (nDOC) to redshift space distortions (RSD); see the discussion after eqs.~\eqref{eq:delta_s_1st} and~\eqref{eq:delta_s_2nd}. In this appendix, we calculate how large these corrections are for the angular power spectrum.

In linear theory, the angular power spectrum including nDOC is
\eq{
C_{\ell} &= \lb \frac{2}{\pi} \rb^2 \int_{\chi_1\chi_2}D_1 D_2 \int k^2 dk \left[
\lb b_{10}j_{\ell}(k\chi_1) - f j''_{\ell}(k\chi_1) - \frac{2f}{k\chi_1} j'_{\ell}(k\chi_1)\rb \right. \non \\
&\hspace{0.3\textwidth} \left. \times \lb b_{10}j_{\ell}(k\chi_2) - f j''_{\ell}(k\chi_2) - \frac{2f}{k\chi_2} j'_{\ell}(k\chi_2)\rb P(k) \right] \, .
}
We separate the equation as $C_\ell = C_\ell^{\mathrm{RSD}} + C_\ell^{\mathrm{nDOC}}$, where
\eq{
C_\ell^{\mathrm{RSD}} = \lb \frac{2}{\pi} \rb^2 \int_{\chi_1\chi_2}D_1 D_2 \int k^2 dk \left[ b_{10}j_{\ell}(k\chi_1) - f j''_{\ell}(k\chi_1) \right]  \left[ b_{10}j_{\ell}(k\chi_1) - f j''_{\ell}(k\chi_2) \right] P(k) 
}
is the commonly used Kaiser RSD part, while $C_\ell^{\mathrm{nDOC}}$ is the part arising from the distant-observer correction. 
\begin{figure}
    \centering
    \includegraphics[width=0.6\linewidth]{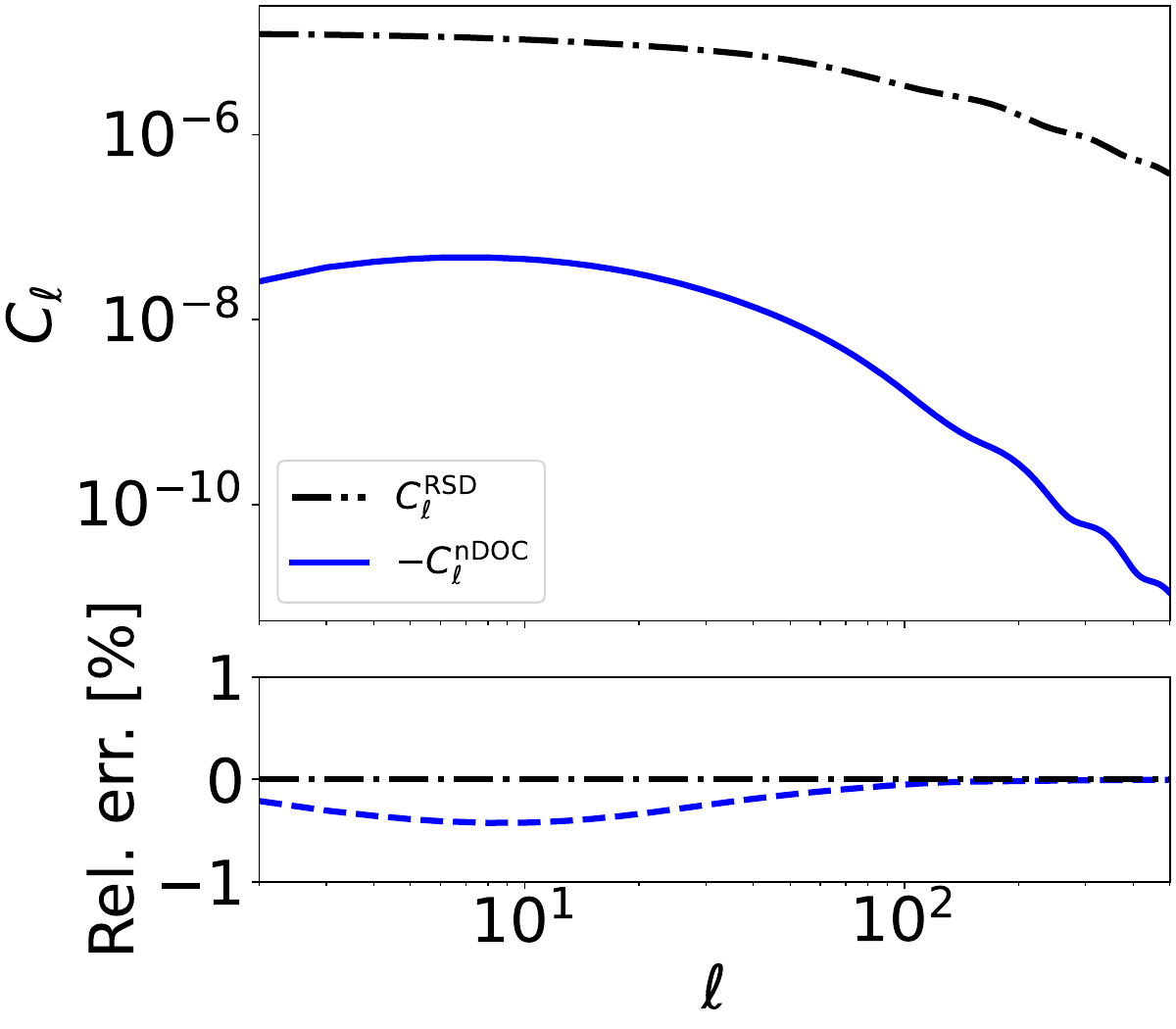}
    \caption{Non-distant observer correction (nDOC) to the angular power spectrum for sources around $z=1.0$ (blue, solid lines), compared to the canonical result ignoring these corrections (black, dot-dashed lines). We set $\sigma_z=0.05$ and $b_{10}=1$.
    The nDOC contributes only up to $0.5\%$ of the canonical RSD spectrum.}
    \label{fig:displace}
\end{figure}

As shown in figure~\ref{fig:displace}, the nDOC RSD term contributes less than $0.5\%$ to the angular power spectrum for sources around $z=1.0$.
Thus, this effect can be ignored given that we are considering redshifts around and above this value.

Although we focused only on the angular power spectrum, we argue that since the tree-level angular bispectra receive contributions from the products of pairs of (unprojected) angular power spectra, the contribution of nDOC to angular bispectra should be negligible as well. Therefore, we do not explicitly evaluate the nDOC terms for the angular bispectra. Note that it is products of unprojected spectra, though, for the bispectrum, but figure~\ref{fig:displace} is for projected angular power spectrum.

\section{Extended Limber approximation for the squeezed limit}\label{append:extended_limber}
\subsection{Real-space case}
\label{subappend:realspace}

For squeezed configurations, we need to be more careful with the Limber approximation. Since one of the multipoles ($\ell_3$, say) is chosen to be soft, the typical fluctuation wavelength $\chi_3/\ell_3$ may not be narrow compared to the width of the window function $W_3(\chi_3)$ and the Limber approximation may not hold. We found that simply making the replacement in eq.~\eqref{eq:spherical_bessel_delta} for all occurrences of the spherical Bessel functions for the hard modes $\ell_1$ and $\ell_2$ is not accurate. A better approximation is obtained by replacing only the $j_{\ell_i}(k_i \chi_i)$ for  $i=1,2$, but not the spherical Bessel functions in the geometrical projection function $\mathcal{D}_{\ell_1\ell_2\ell_3}$.
Thus, choosing $\ell_3$ to be the soft mode, we adopt the following Limber-like approximation for the squeezed configuration:
\eq{
\label{eq:bispec_real_explicit_limber_squeezed}
B^{\rm Lim.}_{\ell_1\ell_2\ell_3} 
&= \lb \frac{2}{\pi} \rb^3 \int_{\chi_1\chi_2\chi_3} \prod_{i=1}^2\lb \lb \frac{\ell_i}{\chi_i}\rb^2  \sqrt{\frac{\pi}{2\ell_i}} \frac{1}{\chi_i} \rb \int k_3^2 dk_3 j_{\ell_3}(k_3 \chi_3) \non \\
&\hspace{1cm} \times \int r^2 dr j_{\ell_1}\lb \frac{\ell_1}{\chi_1} r \rb j_{\ell_2}\lb \frac{\ell_2}{\chi_2} r \rb j_{\ell_3}\lb k_3 r \rb B\lb \frac{\ell_1}{\chi_1}, \frac{\ell_2}{\chi_2}, k_3; z[\chi_1], z[\chi_2], z[\chi_3]\rb \, . 
}

\subsection{Redshift-space}

In redshift space, besides the issues discussed in section~\ref{subappend:realspace},
we also need to include the RSD corrections for the soft mode. First, we consider the SOD term, which dominates on the scales of interest. By choosing $\ell_3$ as the soft mode, we have
\eq{
\label{eq:bispec_SOD_explicit_limber_squeezed}
B_{\ell_1\ell_2\ell_3}^{\rm SOD, Lim.} 
&= \lb \frac{2}{\pi} \rb^3 \int_{\chi_1\chi_2\chi_3}\Bigg[
 D_1D_2 D_3^2 b_{10}(\chi_1) b_{10}(\chi_2)
 \prod_{i=1}^2\lb \lb \frac{\ell_i}{\chi_i}\rb^2  \sqrt{\frac{\pi}{2\ell_i}} \frac{1}{\chi_i} \rb \int k_3^2 dk_3 j_{\ell_3}(k_3 \chi_3) \non \\
&\hspace{0.5cm} \times \int r^2 dr j_{\ell_1}\lb \tfrac{\ell_1}{\chi_1} r \rb j_{\ell_2}\lb \tfrac{\ell_2}{\chi_2} r \rb j_{\ell_3}\lb k_3 r \rb \mathcal{K}_2 \lb \tfrac{\ell_1}{\chi_1}, \tfrac{\ell_2}{\chi_2}, k_3;\;\chi_3 \rb P_L\lb \tfrac{\ell_1}{\chi_1} \rb P_L\lb \tfrac{\ell_2}{\chi_2} \rb  \non \\
& + 2 D_1 D_2^2 D_3 b_{10}(\chi_1) \lb \frac{\ell_1}{\chi_1}\rb^2  \sqrt{\frac{\pi}{2\ell_1}} \frac{1}{\chi_1}
\int k_2^2 dk_2\, j_{\ell_2}\lb k_2 \chi_2\rb \non \\
&\hspace{0.5cm} \times 
\int k_3^2 dk_3\, \lb b_{10} j_{\ell_3}(k_3 \chi_3) - f_3 j''_{\ell_3}(k_3 \chi_3) \rb \non\\
&\hspace{0.5cm} \times \int r^2 dr j_{\ell_1}\lb \tfrac{\ell_1}{\chi_1} r \rb  j_{\ell_2}\lb k_2 r \rb
j_{\ell_3}\lb k_3 r \rb \mathcal{K}_2 \lb \tfrac{\ell_1}{\chi_1}, k_3, k_2;\; \chi_2\rb P_L\lb k_3 \rb P_L\lb \tfrac{\ell_1}{\chi_1} \rb \Bigg] \non \\ & \hspace{10cm} + \text{perm}_{1\leftrightarrow 2}\, .
}
Note that the first term arises from second-order corrections to the soft mode and so is expected to be subdominant. 

Moreover, in the squeezed-limit configuration, we also find numerically that the DV term has a non-negligible contribution on small scales. Thus, we need to include this term in the partial-Limber approximation as 
\eq{
\label{eq:3pt_correlator_DV_limber}
B_{\ell_1\ell_2\ell_3}^{\rm DV, Lim.} 
&= 
-\lb \frac{2}{\pi}\rb^2 \int_{\chi_1\chi_2\chi_3}
\Bigg[  D_1 D_2 D_3^2 f_3 b_{10}(\chi_3)\prod_{i=1}^2\lb \lb \frac{\ell_i}{\chi_i}\rb^2 \sqrt{\frac{\pi}{2\ell_i}} \frac{1}{\chi_i} b_{10}(\chi_i) \rb  \non \\
&\hspace{-0.5cm}\times \lb j_{\ell_1}\lb \tfrac{\ell_1}{\chi_1}\chi_3 \rb j''_{\ell_2}\lb \tfrac{\ell_2}{\chi_2}\chi_3 \rb + \frac{\ell_1}{\ell_2} \frac{\chi_2}{\chi_1} j'_{\ell_1}\lb \tfrac{\ell_1}{\chi_1} \chi_3 \rb j'_{\ell_2}\lb \tfrac{\ell_2}{\chi_2} \chi_3 \rb \rb P_L\lb \tfrac{\ell_1}{\chi_1} \rb P_L\lb \tfrac{\ell_2}{\chi_2} \rb \non \\
&\hspace{-0.5cm} + D_1 D_2^2 D_3 f_2 b_{10}(\chi_2)\lb \frac{\ell_1}{\chi_1}\rb^2  \sqrt{\frac{\pi}{2\ell_1}} \frac{1}{\chi_1} 
\int k_3^2 dk_3\, \bigg( \lb b_{10} j_{\ell_3}(k_3 \chi_3) - f_3 j''_{\ell_3}(k_3 \chi_3) \rb \non\\
&\hspace{-0.85cm} \times \left. \lb j_{\ell_1}\lb k_1 \chi_2 \rb j''_{\ell_3}\lb k_3 \chi_2 \rb + \frac{k_1}{k_3} j'_{\ell_1}\lb k_1\chi_2\rb j'_{\ell_3}(k_3 \chi_2)  +\mathrm{perm}_{1\leftrightarrow 3} \rb \right|_{k_1 = \tfrac{\ell_1}{\chi_1}}P_L\lb \tfrac{\ell_1}{\chi_1} \rb P_L\lb k_3 \rb \bigg) \Bigg] \non \\
&\hspace{10cm} +\mathrm{perm}_{1\leftrightarrow 2} \, .
}

Due to their complexity, these partial-Limber expressions for the squeezed limit are no longer very convenient and simple to evaluate numerically and thus do not provide any practical benefit compared to our full results. However, we do present them in our numerical results in the next subsection for completeness. These numerical results include a Limber version of the (scale-dependent) stochastic term given by
\eq{
B_{\ell_1\ell_2\ell_3}^{\rm Stoch. Lim.} 
&= 2
\int d\chi_3 \frac{W_1(\chi_3) W_2(\chi_3) W_3(\chi_3)}{\chi_3^4} D^2(\chi_3) P_L\lb \frac{\ell_1}{\chi_3}\rb b^2_{10}(\chi_3) P_{\eps \eps_\delta}(\chi_3, \chi_3) +\mathrm{perm}_{1\leftrightarrow 2}  \non\\
&+ 2 \lb \frac{2}{\pi}\rb\int d\chi_1 d\chi_3 \frac{W_1(\chi_3) W_2(\chi_3) W_3(\chi_1)}{\chi_3^2} D(\chi_1) D(\chi_3) \non \\
& \hspace{-1.5cm}\times \int k_1^2 dk_1 P_L(k_1) \left( b_{10}(\chi_1) j_{\ell_3}(k_1 \chi_1) - f(\chi_1) j''_{\ell_3}(k_1 \chi_1) \right) \left( b_{10}(\chi_3) j_{\ell_3}(k_1 \chi_3) - f(\chi_3) j''_{\ell_3}(k_1 \chi_3) \right) \, . 
}

\subsection{Numerical results}

\begin{figure}
\centering
\begin{subfigure}{0.7\textwidth}
    \includegraphics[width=\textwidth]{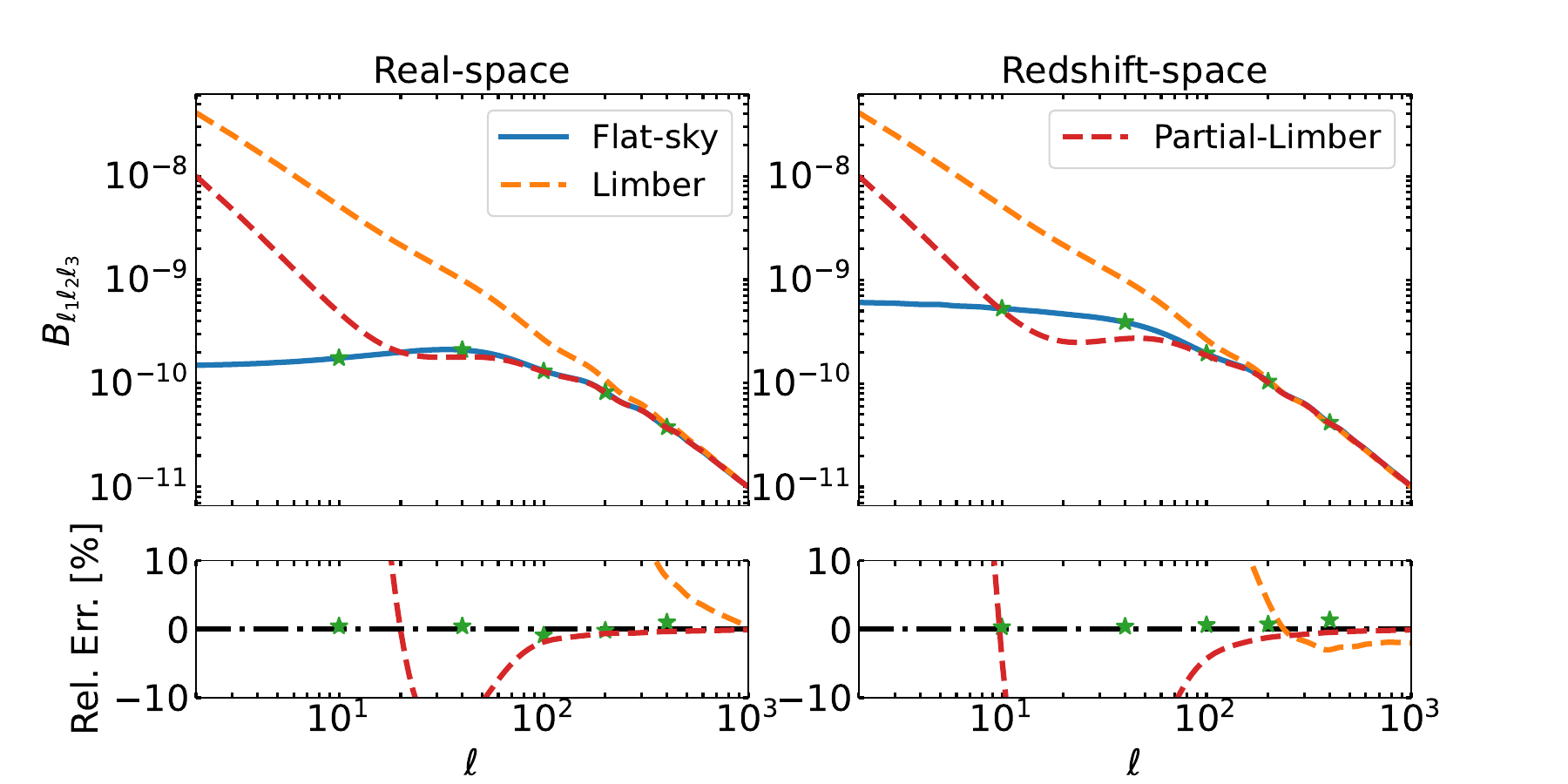}
    \caption{Equal-time}
    \label{fig:first_extended_limber}
\end{subfigure}
\begin{subfigure}{0.7\textwidth} \includegraphics[width=\textwidth]{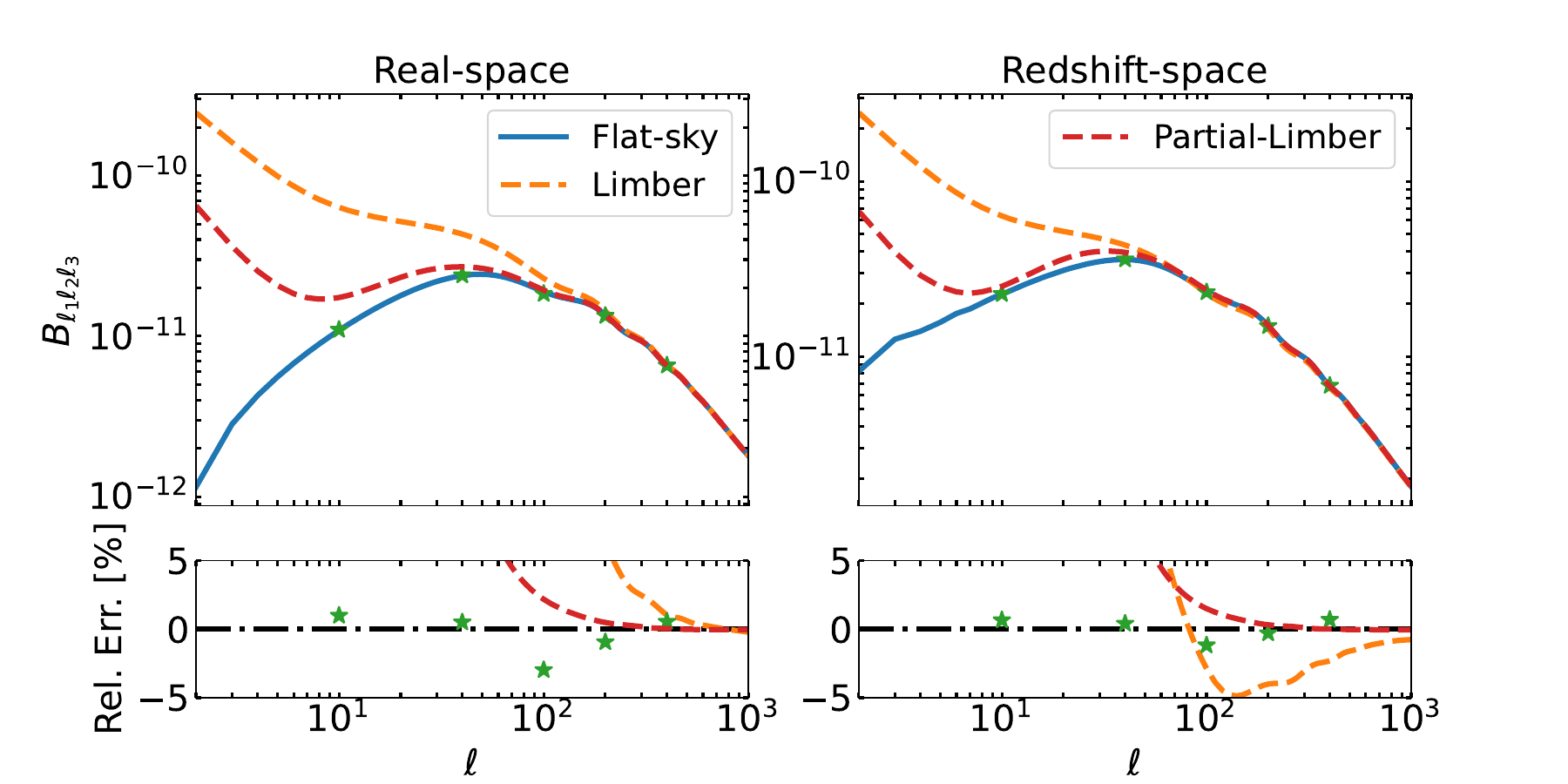}
    \caption{Unequal-time}
    \label{fig:second_extended_limber}
\end{subfigure}
\caption{Comparison of results for the partial-Limber approximation (red dashed lines) and the canonical Limber approximation (orange dashed line), as used in the main text, in the squeezed limit. The flat-sky and (brute-force) full-sky results are shown as blue lines and green stars, respectively. The first row is the equal-time case, as in figure~\ref{fig:equal_time_rsd}, while the second row is unequal-time, as in figure~\ref{fig:unequal_time_rsd}. The results in the left-hand plots do not include RSD while the right-hand plots do. 
}
\label{fig:extended_limber}
\end{figure}

We compare results for our partial-Limber approximation to the canonical Limber approximation for the squeezed configuration in figure~\ref{fig:extended_limber}. We consider equal- and unequal-time cases with the same parameters as in figures~\ref{fig:equal_time_rsd} and~\ref{fig:unequal_time_rsd}, respectively.

We find that in both cases, the partial-Limber approximation extends the agreement with the flat-sky (or full-sky) results to lower multipoles compared to the canonical Limber approximation, albeit at greater computational cost. We also note that the poorer performance of the canonical Limber approximation on small scales in the RSD cases compared to real space is due to the neglect of all RSD corrections.

\bibliographystyle{JHEP}
\bibliography{mybib}

\end{document}